\documentclass[usenatbib,times,onecolumn,usegraphicx,useAMS]{mn2e}
\usepackage{amssymb,url,times,amsmath,fleqn}

\newcommand{\pder}[2]{\ensuremath{\frac{\partial #1}{\partial #2}}}
\def\hterm{{\bf H}}
\def\htensor{H}

\title[MHD in SPH using the Vector Potential]{Smoothed Particle
Magnetohydrodynamics IV - Using the Vector Potential}

\author[Price]{Daniel J. Price\\
Centre for Stellar and Planetary Astrophysics, School of Mathematical Sciences, Monash University, Clayton 3800, Australia
}
\date{Submitted: 27th July 2009 Revised: 14th Sep 2009 Accepted:}
\pagerange{\pageref{firstpage}--\pageref{lastpage}} \pubyear{2009}

\begin{document}
\label{firstpage}
\bibliographystyle{mn2e}
\maketitle

\begin{abstract}
 In this paper we investigate the use of the vector potential as a means of maintaining the divergence constraint in the numerical solution of the equations of Magnetohydrodynamics (MHD) using the Smoothed Particle Hydrodynamics (SPH) method. We derive a self-consistent formulation of the equations of motion using a variational principle that is constrained by the numerical formulation of both the induction equation and the curl operator used to obtain the magnetic field, which guarantees exact and simultaneous conservation of momentum, energy and entropy in the numerical scheme. This leads to a novel formulation of the MHD force term, unique to the vector potential, which differs from previous formulations. We also demonstrate how dissipative terms can be correctly formulated for the vector potential such that the contribution to the entropy is positive definite and the total energy is conserved.

 On a standard suite of numerical tests in one, two and three dimensions we find firstly that the consistent formulation of the vector potential equations is unstable to the well-known SPH tensile instability, even more so than in the standard Smoothed Particle Magnetohydrodynamics (SPMHD) formulation where the magnetic field is evolved directly. Furthermore we find that, whilst a hybrid approach based on the vector potential evolution equation coupled with a standard force term gives good results for one and two dimensional problems (where $dA_{z}/dt = 0$), such an approach suffers from numerical instability in three dimensions related to the unconstrained evolution of vector potential components. We conclude that use of the vector potential is not a viable approach for Smoothed Particle Magnetohydrodynamics.
\end{abstract}


\section{Introduction}
\label{sec:intro}
 Magnetic fields are important at some level in almost every area of astrophysics and their effects are commonly accounted for in numerical simulations by solving the equations of magnetohydrodynamics (MHD). The solution of the MHD equations throws up a number of challenges for any numerical method, most notably because of the divergence-free (``no monopoles'') constraint on the magnetic field, which does not appear explicitly in the MHD equations but rather as an initial condition which, if satisfied initially, should remain satisfied for all time (\citealt{toth00}; Price \& Monaghan 2005).
 
 Attempts to solve the equations of MHD using the Smoothed Particle Hydrodynamics (SPH) method (for recent reviews see \citealt{price04,monaghan05}) have a long and somewhat tortured history, beginning with one of the earliest SPH papers by \citet{gm77}, though not seriously developed until \citet{pm85}. The latter authors discovered that the equations of Smoothed Particle Magnetohydrodynamics (SPMHD) contained a catastrophic numerical instability (now known as the ``tensile instability'', \citealt{monaghan00}) when written in a form that conserved momentum exactly, albeit only in a certain regime (magnetic pressure greater than gas pressure). Despite detailed investigation by \citet{morrisphd} and \citet{meglicki95}, these problems meant that, apart from a few isolated applications \citep[e.g.][]{dbl99}, SPMHD did not find widespread adoption.
 
  More recently, progress has been made on a number of fronts, in particular with regards to formulating dissipative terms in order to handle MHD shocks (Price \& Monaghan 2004a, hereafter \citealt{pm04a}) and in formulating the SPMHD equations self-consistently from a variational principle (Price \& Monaghan 2004b, hereafter \citealt{pm04b}). In \citet{pm04b} we have furthermore derived the equations of motion accounting for terms related to the smoothing length gradients which are necessary for exact simultaneous conservation of both energy and entropy. There is also a reasonable consensus on good approaches to removing the tensile instability in SPMHD, using either formulations proposed by \citet{morrisphd} or by \citet*{bot01} which forgo the exact conservation of momentum slightly in order to attain numerical stability, the latter method simply by subtracting off the term in the force equation which is proportional to the non-zero divergence. Application of the \citet{monaghan00} fix for the tensile instability \citep{pm04a} turned out to be unsuccessful in compressible flows \citep{price04}.
  
   In Price \& Monaghan (2005) (hereafter \citealt{pm05}), an extensive investigation was made into methods for addressing the divergence constraint in the context of SPMHD, using either projection methods or hyperbolic cleaning schemes. However, whilst such schemes could be made to give good results on test problems (which can also be successfully run without any form of divergence cleaning), it was found that they performed poorly for ``real'' applications such as that of star formation where length and time scales can change by many orders of magnitude \citep{pb07}. For this reason our attention shifted to an earlier formulation adopted by \citet{pm85} (though later discarded due to poor accuracy when calculated with the spatially constant smoothing lengths and simple gradient operators used by these authors) whereby the magnetic field was formulated in terms of the so-called ``Euler potentials'' \citep{stern70,stern76} (referred to as ``Clebsch variables'' by \citealt{pm85}), $\alpha_{E}$ and $\beta_{E}$, where the magnetic field is expressed as
\begin{equation}
{\bf B} = \nabla\alpha_{E} \times \nabla\beta_{E}.
\label{eq:Beuler}
\end{equation}
The advantages are twofold - the first is that the divergence constraint is satisfied by construction (that is, taking the divergence of (\ref{eq:Beuler}) gives zero). The second is that 
the induction equation in ideal MHD written in terms of the Euler potentials takes a particularly simple form, namely
\begin{equation}
\frac{d\alpha_{E}}{dt} = 0, \hspace{5mm} \frac{d\beta_{E}}{dt} = 0,
\label{eq:eulerevol}
\end{equation}
which corresponds physically to the advection of magnetic field lines by Lagrangian particles \citep{stern66}.

 Using the Euler potentials formulation together with the dissipative terms and force equation proposed in \citet{pm05}, has meant that the SPMHD algorithm has been successfully applied to a number of ``real-world'' problems, including neutron star mergers \citep{pr06}\footnote{Although it later turned out that there were problems in applying the Euler potentials formulation in the context of neutron star mergers.}, star formation \citep{pb07,pb08} and the dynamics of spiral galaxies \citep{dp08,kotarbaetal09}. Furthermore this has led to the development of at least two ``magnetic-SPH'' codes \citep{rp07,ds08}, the latter adding MHD to the widely used GADGET code for cosmological simulations \citep{springel05}.

 However there are also important limitations to the Euler potentials approach, namely that the magnetic helicity
\begin{equation}
 \int {\bf A}\cdot {\bf B}\phantom{d}{\rm dV},
\end{equation}
where ${\bf A}$ is the magnetic vector potential and ${\bf B}$ is the magnetic field, is constrained to be zero by a simple consequence of the fact that ${\bf A} = \alpha_{E} \nabla\beta_{E}$ (equivalently ${\bf A} = -\beta_{E} \nabla\alpha_{E}$ by a change of gauge) is exactly perpendicular to ${\bf B}$, implying that ${\bf A}\cdot{\bf B} = 0$. In practise this means that firstly it is simply not possible to represent certain fields using Euler potentials, as they would become double valued. For example, one can easily represent either a toriodal or a poloidal field using Euler potentials \citep{stern76} but not a combination of both\footnote{One possible way around this restriction is to note that one can construct any given magnetic field by a linear combination of ${\bf B}$'s, where each ${\bf B}$ is determined by a separate set of Euler potentials, as done for example by \citet{yl08}.}. The corollary is also that such complex fields cannot be created during the simulation. A better way of understanding this limitation in practise is to recognise that equation (\ref{eq:eulerevol}), since there is no time evolution of the potentials on the particles, represents a mapping of the magnetic field on the initial particle configuration at $t=0$ to a new arrangement on the particle configuration at some later time $t$, and can change the geometry of the field only insofar as a one-to-one mapping from the initial to the final particle distribution exists (i.e., the field can only be followed for $\lesssim$ 1 dynamical time). Thus important physical processes such as winding up of fields by differential rotation are largely missed by the Euler potentials formulation (see \citealt{brandenburg09} for some examples) and this motivates us to consider a more general approach.
 
  In this fourth paper we examine the question of whether a formulation of SPMHD based on the magnetic vector potential ${\bf A}$ can resolve the difficulties associated with the Euler potentials formulation whilst at the same time maintaining the divergence constraint on the magnetic field. In the vector potential formulation the magnetic field is given by
\begin{equation}
{\bf B} = {\bf B}_{int} + {\bf B}_{ext} = \nabla\times {\bf A} + {\bf B}_{ext},
\label{eq:BcurlA}
\end{equation}
where ${\bf B}_{int}\equiv\nabla\times{\bf A}$ is the magnetic field due to internal currents i.e., those within the computation domain, and here we assume that ${\bf B}_{ext}$ is a time-independent externally applied field. The time evolution of ${\bf A}$ can be derived from the induction equation for the magnetic field
\begin{equation}
\pder{\bf B}{t} = \nabla \times ({\bf v} \times {\bf B}) - \nabla \times (\eta {\bf J}),
\end{equation}
where $\eta$ is the magnetic resistivity ($\eta \equiv 1/(\sigma \mu_{0}$) where $\sigma$ is the conductivity and $\mu_{0}$ is the permeability of free space)  and ${\bf J} = \nabla\times{\bf B}/\mu_{0}$ is the current density. `Uncurling' this equation, we have
\begin{equation}
\pder{\bf A}{t} = {\bf v} \times {\bf B} - \eta{\bf J} + \nabla \phi,
\end{equation}
where $\phi$ is an arbitrary scalar representing the freedom to choose a gauge. In Lagrangian form, and expanding ${\bf B}$ in terms of internal ${\bf B}_{int} = \nabla\times{\bf A}$ and external ${\bf B}_{ext}$  (i.e., produced by currents outside the simulation domain) components this is given by
\begin{equation}
\frac{d{\bf A}}{dt} = {\bf v} \times \nabla \times {\bf A} + ({\bf v}\cdot\nabla) {\bf A} + {\bf v} \times {\bf B}_{ext} -\eta{\bf J} + \nabla \phi.
\label{eq:indA}
\end{equation}
 
 Part of the difficulty in assessing the usefulness or otherwise of the vector potential is that differences between the Euler potentials and the vector potential \emph{only} occur in three dimensions, since for one and two dimensional problems the two formulations are identical \citep[e.g.][]{rp07}, since in two dimensions the vector potential has only one component $A_{z} \equiv \alpha_{E}$ which evolves according to $dA_{z}/dt = 0$ (for the Euler potentials $\nabla\beta \equiv {\bf z}$ in 2D). However a rigourous formulation of the SPMHD equations of motion has not been derived for either the vector or Euler potentials which has relevance to both formulations even in one and two dimensions.
 
  The approach we take is to use a Lagrangian variational principle (similar to the approach taken in \citealt{pm04b} for the standard case) in order to derive the equations of motion in a manner that is constrained by the exact numerical formulation of both (\ref{eq:BcurlA}) and the numerical representation of the induction equation for the vector potential (\S\ref{sec:Lagrangian}). This means that exact conservation of \emph{all} physical quantities (linear and angular momentum and energy) is guaranteed in the resultant numerical equations provided that the appropriate symmetries (i.e., invariance to translations, rotations and time, respectively) are present in the Lagrangian and the equations used to constrain it. We find that this very powerful approach leads to a novel formulation of the force term which is already different to previous SPMHD formulations of the MHD force in one and two dimensions and indeed conserves momentum and energy exactly. We demonstrate that these symmetries are also respected in three dimensions provided an appropriate gauge choice is made in the $d{\bf A}/dt$ equation in order that it is Galilean invariant.
  
  Secondly, in \S\ref{sec:diss} we show how dissipative terms should be constructed for vector potential SPMHD in order that total energy is conserved and that the second law of thermodynamics is obeyed, i.e., a positive definite contribution to the entropy results. These terms, which are derived independently of the equations of motion, differ from previous formulations of dissipative terms that have been used for the vector/Euler potentials in SPMHD.

  Finally, we examine the new vector potential force formulation and the dissipative terms on the suite of one and two dimensional test problems (\S\ref{sec:tests}). Whilst the hope was that by constructing the SPMHD equations such that the divergence-free constraint was inbuilt, instabilities would not appear in the equations. However it turns out that the consistent formulation of the vector potential force has similar  -- in fact, much worse -- problems with the tensile instability than even the standard conservative SPMHD force. Whilst we have managed to obtain reasonable results on a range of numerical tests with the consistent vector potential equations of motion, we find that a better approach is to use the vector potential in conjunction with a stable but non-conservative force such as those employed in \citet{pm05} and \citet{bot01}. The main practical improvement in this paper is therefore in the formulation of the dissipative terms.
  
\section{A consistent formulation of SPMHD using the vector potential}
\label{sec:consistent}

\subsection{Variational Principle}
\label{sec:Lagrangian}
 We start from the Lagrangian for MHD, which in continuum form is given by
(e.g. \citealt{newcomb62})
\begin{equation}
L = \int \left(\frac{1}{2}\rho \mathbf{v}^2 -\rho u-\frac{1}{2\mu_0}B^2\right)
\mathrm{dV},
\end{equation}
which is simply the kinetic minus the {\bf thermal} and magnetic energies. The SPH
Lagrangian is obtained, following \citet{pm04b} and \citet{mp01} by replacing the integral by a summation and the mass element $\rho\rm{dV}$ by the mass per SPH particle $m$, giving
\begin{equation}
L_{sph} = \sum_{b=1}^{N} m_{b} \left[\frac{1}{2}\mathbf{v}_b^2 - u_b(\rho_b,s_b) -\frac{1}{2\mu_0}
\frac{B_b^2}{\rho_b}\right].
\label{eq:lagrangian}
\end{equation}
where ${\bf v}\equiv \dot{\bf x}$ is the velocity,  $\rho$ is the density, $u$ is the thermal energy per unit mass (in general a function of both density $\rho$ and entropy $s$) and ${\bf B}$ is the magnetic field.

 The equations of motion can be derived using the Euler-Lagrange equations provided that all variables appearing in the Lagrangian can be expressed as a function of the particle coordinates and velocities \citep[e.g.][]{mp01,pm07}. Whilst for hydrodynamics the density can be written directly as a function of the particle coordinates via the SPH density sum (which is an exact solution of the continuity equation), for MHD the magnetic field ${\bf B}$ (in this case the vector potential ${\bf A}$) can only be written as a function of the \emph{change} in particle coordinates (i.e., we do not have an exact solution to the induction equation). In this case (as in \citealt{pm04b}) we can derive the equations of motion by perturbing the Lagrangian and specifying that the change in action is zero, i.e.,
\begin{equation}
\delta S = \int \delta L {\rm  dt} = 0.
\label{eq:deltaS}
\end{equation}
where the variation $\delta L$ is with respect to a small change in the particle coordinates $\delta{\bf x}$. Importantly conservation properties (e.g. momentum conservation) in this case will only follow provided that the respective symmetries (e.g. invariance to translation) are preserved in the numerical representation of the perturbation.

 Perturbing the Lagrangian (Equation \ref{eq:lagrangian}) with respect to a change in the position of particle $a$, i.e., $\delta {\bf x}_{a}$, we have
\begin{equation}
\delta L = m_a {\bf v}_a\cdot\delta{\bf v}_a- \sum_{b} m_{b}\left[\left.\pder{u_b}{\rho_b}\right\vert_s\delta\rho_b - \frac{3}{2\mu_0} \left(\frac{B_b}{\rho_b}\right)^2\delta\rho_b + \frac{1}{\mu_0}
\frac{{\bf B}_b}{\rho_{b}^{2}}\cdot\delta\left(\rho_{b}{\bf B}_b\right)\right],
\label{eq:deltaL}
\end{equation}
where we have expressed the perturbation for the magnetic field in terms of $\delta (\rho {\bf B})$ for reasons that will become clear. The equations of motion are obtained by using (\ref{eq:deltaL}) in (\ref{eq:deltaS}) and integrating the velocity term by parts with respect to time, i.e.,
\begin{equation}
\int m_{a} {\bf v}_{a} \cdot \delta \left(\frac{d{\bf x}_{a}}{dt}\right){\rm dt} = \left[ m_{a} {\bf v}_{a}\cdot \delta{\bf x}_{a}\right]^{t}_{0} - \int m_{a} \frac{d{\bf v}_{a}}{dt}\cdot\delta{\bf x}_{a}{\rm dt},
\end{equation}
giving (in tensor notation to avoid confusion of indices)
\begin{equation}
\int\left\{-m_a \frac{dv^{i}_a}{dt} - \sum_{b} m_{b}\left[\frac{P_{b}}{\rho_{b}^{2}}\frac{\delta\rho_b}{\delta x^{i}_{a}} - \frac{3}{2\mu_0} \left(\frac{B_b}{\rho_b}\right)^2\frac{\delta\rho_b}{\delta x^{i}_{a}} + \frac{1}{\mu_0}
\frac{B^{j}_b}{\rho_{b}^{2}}\frac{\delta\left(\rho_{b}B^{j}_b\right)}{\delta x^{i}_{a}}\right] \right\}\delta x^{i}_a \mathrm{dt} = 0,
\label{eq:deltaLexpanded}
\end{equation}
where we have used the first law of thermodynamics to write $\partial u/\partial \rho\vert_{s} = P/\rho^{2}$ (see \citealt{pm04b}).

 What remains is to express, as SPH summations over neighbours, the perturbations $\delta \rho_{b}$ and $\delta (\rho_{b} {\bf B}_{b})$ taken with respect to $\delta {\bf x}_{a}$. The derivation here is more complicated than that presented by \citet{pm04b} because in this case we must express $\delta {\bf B}$ in terms of $\delta {\bf A}$ (via the SPH expression of Equation \ref{eq:BcurlA}) and in turn, $\delta {\bf A}$ in terms of $\delta {\bf x}$ (via the SPH version of the induction equation for the vector potential). We thus formulate the SPH expression of each of these equations in \S\ref{sec:sph}, below, with the corresponding perturbations presented in \S\ref{sec:perturb}. The equations of motion are derived in \S\ref{sec:equationsofmotion}.

\subsection{SPH formulation}
\label{sec:sph}
\subsubsection{Density sum}
\label{sec:densitysum}
 We base our SPMHD formulation for the vector potential on the variable smoothing length formulation of SPH presented by \citet{pm04b,pm07} (see also \citealt{monaghan02,monaghan05}). The density $\rho$ is calculated on particle $a$ from neighbouring particles via the summation
\begin{equation}
\rho_{a} = \sum_{b} m_{b} W(\vert {\bf r}_{a} - {\bf r}_{b}\vert , h_{a}),
\label{eq:rhosum}
\end{equation}
where $W$ is the SPH kernel function, details of which are given in Appendix~\ref{sec:kernel}. Key to the variable smoothing length formulation is whilst the kernel function depends on the smoothing length $h$, $h$ itself is defined as a function of the particle positions, expressed most conveniently as a function of the density sum itself, via the relation
\begin{equation}
h_a = \eta \left(\frac{m_{a}}{\rho_a} \right)^{1/\nu},
\label{eq:hrho}
\end{equation}
with derivatives
\begin{equation}
\pder{h_a}{\rho_{a}} = -\frac{h_a}{\nu\rho_a}, \hspace{1cm} \pder{^{2} h_a}{\rho_{a}^{2}} = \frac{h_a}{\rho_a^{2}} \left( \frac{\nu + 1}{\nu^{2}} \right) ,
\label{eq:dhdrho}
\end{equation}
where $\nu$ is the number of spatial dimensions and $\eta$ is a dimensionless constant specifying the smoothing length in terms of the mean inter-particle spacing. Equation (\ref{eq:hrho}) in turn determines the ``number of neighbours'' in the SPH calculation. Unless otherwise indicated we use $\eta = 1.2$ throughout this paper, corresponding to $\sim 58$ neighbours in 3D for kernels with a compact support radius of $2h$.

 The density summation is therefore a non-linear equation for both $\rho$ and $h$ that we solve using a Newton-Raphson scheme as described in detail in \citet{pm07}. Enforcement of the relationship between $h$ and $\rho$ is a necessary requirement for energy conservation, since hereafter we will assume that the smoothing length is differentiable with respect to particle position.

\subsubsection{$\nabla\times {\bf A}$}
\label{sec:BcurlA}
 In principle we have a number of choices for the numerical formulation of both equation (\ref{eq:BcurlA}) and the induction equation. In practise these choices are constrained by the requirement that symmetries in the Lagrangian are preserved by the constraint equations. For example, in calculating (\ref{eq:BcurlA}) we can in principle use any of the SPH curl operators as discussed e.g. in \citet{price04}. The basic operation for a curl in SPH is given by
\begin{equation}
(\nabla\times{\bf A})_{a} = -\sum_{b} \frac{m_{b}}{\rho_{b}} {\bf A_{b}} \times \nabla_{a} W_{ab},
\end{equation}
where $W_{ab} \equiv W(\vert {\bf r}_{a} - {\bf r}_{b}\vert , h)$ is the SPH kernel. In principle the kernel used for the curl does not have to be the same kernel as used in the density summation, though in this paper we assume that this is the case. Since the time evolution of (and thus the perturbation to) ${\bf A}$ is not in itself invariant to translations in {\bf v} (for example in the case of an external field, see below), we require a curl operator for $\nabla\times{\bf A}$ such that ${\bf B}$ is invariant to the addition of an arbitrary constant to $\bf A$ such that the Lagrangian is Galilean invariant and therefore that the resultant force will conserve momentum. In the variable smoothing length formulation of SPH, this can be achieved using
\begin{equation}
{\bf B}_{a} = (\nabla\times{\bf A})_{a} + {\bf B}_{ext} = \frac{1}{\Omega_{a}\rho_{a}} \sum_{b} m_{b} ({\bf A}_{a} - {\bf A}_{b}) \times \nabla_{a} W_{ab} (h_{a}) + {\bf B}_{ext},
\label{eq:curlA}
\end{equation}
where $\Omega_{a}$ is a normalisation factor related to the smoothing length gradient, given by
\begin{equation}
\Omega_{a} = 1 - \pder{h_{a}}{\rho_{a}}\sum_{b} m_{b} \pder{W_{ab}}{h_{a}},
\label{eq:omega}
\end{equation}
which can be calculated alongside (\ref{eq:curlA}) and the density sum (\ref{eq:rhosum}). Note that we could equally have chosen the alternative SPH curl formulation which has a $1/\rho_{b}$ inside the summation instead of the above (see e.g. \citealt{price04}) --- equation (\ref{eq:curlA}) has the advantage that it does not require prior computation of the density (which involves a summation over the particles) and depends only on the particle's own smoothing length (i.e., $h_{a}$). The symmetric formulation of the curl, i.e.,
\begin{equation}
(\nabla\times {\bf A})_{a} = -\rho_{a} \sum_{b} m_{b} \left[ \frac{{\bf A}_{a} \times \nabla_{a} W_{ab} (h_{a})}{\Omega_{a} \rho_{a}^{2}}  + \frac{{\bf A}_{b} \times \nabla_{a} W_{ab} (h_{b})}{\Omega_{b}\rho_{b}^{2}} \right],
\label{eq:symmetriccurl}
\end{equation}
is ruled out by the requirement that $\bf B$ be invariant to the addition of an arbitrary constant to $\bf A$, as discussed above.

\subsubsection{Time evolution of {\bf A}}
\label{sec:dAdt}
 For the induction equation we have freedom to choose not only the SPH formulation of the derivatives in Equation (\ref{eq:indA}) but also to choose an appropriate Gauge for the evolution of the vector potential (i.e., a choice of $\phi$ in Equation \ref{eq:indA}). Again here, our choice is constrained by physical requirements. Most importantly, in three dimensions in order to obtain momentum conservation in the equations of motion, we require that the induction equation is invariant to translations (i.e., is Galilean invariant). A gauge choice which achieves this, first suggested to us by Axel Brandenburg (2007, private communication) is to choose $\phi = {\bf v}\cdot{\bf A}$, which leads to the induction equation in the form
\begin{equation}
\frac{d{\bf A}}{dt} = - {\bf A}\times (\nabla\times {\bf v}) - ({\bf A}\cdot\nabla){\bf v} + {\bf v}\times{\bf B}_{ext} - \eta{\bf J}.
\label{eq:dAdt}
\end{equation}
It turns out that there are other good reasons for this choice of gauge - most notably that this in fact represents the correct low-speed ($v << c$) and magnetically dominated ($E << cB$) limit for electromagnetism \citep{dmr07}.  Written in tensor notation, equation (\ref{eq:dAdt}) can be expressed by
\begin{equation}
\frac{dA_{i}}{dt} = -A_{j} \pder{v^{j}}{x^{i}} + \epsilon_{ijk} v^{j} B_{ext}^{k} - \eta J_{i},
\label{eq:dAdttensor}
\end{equation}
where $\epsilon_{ijk}$ is the Levi-Civita permutation tensor and repeated indices imply a summation. Compare this with a ``naive'' gauge choice $\nabla\phi = 0$, which gives (from equation \ref{eq:indA})
\begin{equation}
\frac{dA_{i}}{dt} = v^{j} \pder{A_{j}}{x^{i}} + \epsilon_{ijk} v^{j} B_{ext}^{k} - \eta J_{i}.
\label{eq:standardgauge}
\end{equation}

It is worth commenting that both (\ref{eq:indA}) and (\ref{eq:dAdt}) are significantly more complicated than the evolution equations for the Euler potentials in 3D (\ref{eq:eulerevol}). The number of derivatives that require numerical evaluation for a Lagrangian code is also one more than would be required in an Eulerian scheme, since here we must compute not only the Eulerian term but also a ``reverse advection'' term to obtain the Lagrangian time derivative on the left hand side. Since advection terms are generally the source of the most difficulty in Eulerian codes -- the main advantage of SPH is that these terms are not present -- the accuracy with which these derivatives can be computed in practise on SPH particles is a concern. Whilst in the Galilean invariant gauge (\ref{eq:dAdt}) the ``reverse advection'' term becomes a derivative of ${\bf v}$ rather than ${\bf A}$, the same concerns apply.

 With regards to the SPH formulation of equation (\ref{eq:dAdttensor}) we face a similar choice of SPH operators for computing both the curl and gradient terms as in \S\ref{sec:BcurlA}. In this case we are again constrained by the physical requirement that the SPH expression of (\ref{eq:dAdttensor}) should be invariant to the addition of an arbitrary constant to ${\bf v}$, which is achieved using a similar operator to that used for the curl of $\bf A$. Neglecting dissipative terms (discussed separately in \S\ref{sec:diss}), we have
\begin{equation}
\frac{dA^{a}_{i}}{dt} = \frac{A^{a}_{j}}{\Omega_{a}\rho_{a}} \sum_{b} m_{b} (v^{j}_{a} - v^{j}_{b}) \pder{W_{ab} (h_{a})}{x^{i}_{a}} + \epsilon_{ijk} v^{j}_{a} B_{ext,a}^{k},
\label{eq:dAdtSPH}
\end{equation}
where in the above and throughout this paper, we adopt the convention that $a,b,c,d$ refer to particle labels whilst $i,j,k,l,m$ and $n$ refer to vector/tensor components. Again, in principle we could also choose the form with $1/\rho_{b}$ inside the summation rather than the above. As in \S\ref{sec:BcurlA}, we choose the above expression because it does not require prior knowledge of the density to compute and can therefore be computed efficiently alongside the density summation.

\subsection{Hybrid approach}
\label{sec:hybrid}
 At this point a hybrid approach would be to compute the time evolution of the vector potential using equation (\ref{eq:dAdtSPH}), calculate a magnetic field $\bf B$ using (\ref{eq:curlA}) and then simply use this $\bf B$ in the equations of motion using (any of) the usual SPH expression(s) for the Lorentz force \citep[e.g.][]{pm04b,pm05}, for example the \citet{morrisphd} formulation:
\begin{equation}
\frac{dv_{a}^{i}}{dt} = - \sum_b m_b\left[\frac{P_a + \frac12 B^2_a/\mu_0}{\Omega_{a}\rho_a^2}\pder{W_{ab}(h_{a})}{x^i} + \frac{P_b + \frac12
B^2_b/\mu_0}{\Omega_{b}\rho_b^2}\pder{W_{ab}(h_{b})}{x^i}\right]  + \frac{1}{\mu_0}\sum_b m_b
\frac{(B_iB_j)_b - (B_iB_j)_a}{\rho_a\rho_b}\overline{\pder{W_{ab}}{x^{j}}},
\label{eq:morrisforce}
\end{equation}
where
\begin{equation}
\overline{\pder{W_{ab}}{x^{j}}} = \frac12 \left[ \pder{W_{ab}(h_{a})}{x^j} + \pder{W_{ab}(h_{b})}{x^j}\right].
\end{equation}
Alternatively one can use the stable MHD force formulated by \citet{bot04}, that is, where the source term $\frac12{\bf B}(\nabla\cdot{\bf B})$ is subtracted from the conservative force, giving
\begin{equation}
\frac{dv_{a}^{i}}{dt} = \sum_b m_b\left[\frac{M^{ij}_{a}}{\Omega_{a}\rho_a^2}\pder{W_{ab}(h_{a})}{x^j} + \frac{M^{ij}_b}{\Omega_{b}\rho_b^2}\pder{W_{ab}(h_{b})}{x^j}\right]  + \frac12 \frac{B^{i}_{a}}{\mu_0}\sum_{b} m_{b} \left[ \frac{B^{j}_{a}}{\Omega_{a}\rho_{a}^{2}}\pder{W_{ab}(h_{a})}{x^{j}_{a}}  + \frac{B^{j}_{b}}{\Omega_{b}\rho_{b}^{2}}\pder{W_{ab}(h_{b})}{x^{j}_{a}}\right].
\label{eq:botforce}
\end{equation}
where
\begin{equation}
M^{ij} =  -P \delta^{ij} + \frac{1}{\mu_{0}}\left(B^{i}B^{j} - \frac12 B^{2}  \delta^{ij}\right).
\end{equation}

  Indeed this is exactly the approach taken using the Euler potentials by e.g. \citet{pb07,pb08}. The flaw in this methodology is that, since the equations of motion are not derived with the constraint of the numerical formulation of the induction equation, there is \emph{no guarantee} that total energy will be conserved\footnote{An important aside with respect to Eulerian codes is due here. Whilst in principle it is possible to enforce total energy conservation in \emph{any} SPH scheme by simply evolving the total energy equation instead of the internal energy equation, if the system is not Hamiltonian all this does is push the errors to another quantity (for example the entropy in the case of hydrodynamics) and in practise would simply lead to negative pressures where `total conservation' is violated. The very power of a Hamiltonian formulation of SPH is that \emph{exact} and \emph{simultaneous} conservation of \emph{all} physical quantities is achieved (i.e., with zero dissipation) something which is never possible in a grid-based scheme. The caveat is that dissipation terms are then explicitly \emph{added} to the SPH scheme in order to capture shocks and other discontinuities, the crudeness of which in practise often makes the scheme \emph{more} dissipative than its grid-based counterpart. However there is no \emph{intrinsic} dissipation in SPH.} (and indeed, using either the Morris force or the \citet{bot04} approach, total momentum conservation is not guaranteed either, though the errors are quite small even for shock-type problems, see \citealt{price04}). In this context total energy conservation means that
\begin{equation}
\frac{dE}{dt} = \frac{d}{dt} \sum_{b} m_{b} \left(\frac12 v_{b}^{2} + u_{b} + \frac12 \frac{B_{b}^{2}}{\mu_{0}\rho_{b}} \right) = \sum_{b} m_{b} \left( {\bf v}_{b}\cdot \frac{d{\bf v}_{b}}{dt} + \frac{du_{b}}{dt} -\frac12 \frac{B_{b}^{2}}{\rho_{b}^{2}}\frac{d\rho_{b}}{dt}+ \frac{{\bf B}_{b}}{\mu_{0}\rho_{b}}\cdot \frac{d{\bf B}_{b}}{dt} \right) = 0,
\label{eq:dEdt}
\end{equation}
where $d{\bf B}/dt$ is the time derivative of (\ref{eq:curlA}) which in turn involves the time derivative of $\bf A$ and thus our induction equation (\ref{eq:dAdtSPH}) (we derive the expression for $d{\bf B}/dt$ in Appendix \ref{sec:deltaBrho}). What is required is that the $d{\bf v}/dt$ term in the above is consistent with the $d{\bf B}/dt$ term resulting from the vector potential evolution. Needless to say, guaranteeing the conservation of energy in a vector potential approach is thus a complicated business, and one which is best achieved by following a variational approach.
 
 Whilst the hybrid approach works reasonably well for the Euler potentials (where the time evolution is zero according to equation \ref{eq:eulerevol}), for the more complicated evolution of the vector potential (Equation \ref{eq:dAdtSPH}), using the induction equation to derive and thus constrain the MHD force term is more important. Furthermore, as we show below, this leads to a novel formulation of the Lorentz force in SPH which has not previously been considered and which has a number of interesting properties. In fact it should be possible to derive the corresponding formulation for the Euler potentials also, however we defer this to a future work. We compare the hybrid approach described above to the consistent vector potential formulation described below in the numerical tests presented in \S\ref{sec:tests}.

\subsection{Variational formulation}
\label{sec:variational}

\subsubsection{Perturbations}
\label{sec:perturb}
 In order to derive the equations of motion from (\ref{eq:deltaS})-(\ref{eq:deltaL}) it remains to express the perturbations $\delta\rho_{b}$ and $\delta(\rho_{b}{\bf B}_{b})$ in terms of the change in particle positions $\delta {\bf x}_{a}$. The change in density is obtained by a perturbation of the density summation, giving \citep{pm04b}
\begin{equation}
\delta \rho_{b} = \frac{1}{\Omega_{b}}\sum_{c} m_{c} (\delta {\bf x}_{b} - \delta {\bf x}_{c}) \cdot \nabla_b W_{bc}(h_{b}),
\label{eq:deltarhovec}
\end{equation}
which, when taken with respect to particle $a$, gives
\begin{equation}
\frac{\delta \rho_{b}}{\delta {\bf x}_{a}} = \frac{1}{\Omega_{b}}\sum_{c} m_{c} (\delta_{ba} - \delta_{ca}) \nabla_b W_{bc}(h_{b}),
\label{eq:deltarho}
\end{equation}
where $\delta_{ba}$ is the Kronecker delta referring to the particle labels.

 The derivation of the Lagrangian perturbation for the magnetic field from (\ref{eq:curlA}) is a little more complicated and is given in Appendix \ref{sec:deltaBrho}. The result, including all terms relating to gradients in the smoothing length, is given by
\begin{eqnarray}
\delta (\rho_{b} {\bf B}_{b}) &= & \frac{1}{\Omega_{b}}\sum_{c} m_{c} ({\bf A}_{b} - {\bf A}_{c}) \times [(\delta {\bf x}_{b} - \delta {\bf x}_{c})\cdot\nabla] \nabla_{b} W_{bc} (h_{b}) \nonumber \\
& + & \frac{1}{\Omega_{b}} \sum_{c} m_{c} \left(\delta{\bf A}_{b} - \delta{\bf A}_{c} \right) \times \nabla_{b} W_{bc} (h_{b}) + {\bf B}_{ext}\delta\rho_{b} \nonumber \\
& + & \left[ \hterm_{b} + \frac{{\bf B}_{b,int}}{\Omega_{b}} \zeta_{b} \right] \delta\rho_{b} + \frac{{\bf B}_{b,int}\rho_{b}}{\Omega_{b}}\pder{h_{b}}{\rho_{b}} \sum_{c} m_{c} \left[(\delta {\bf x}_{b} - \delta {\bf x}_{c})\cdot\nabla_{b}\right] \pder{W_{bc}(h_{b})}{h_{b}},
\label{eq:deltarhoB}
\end{eqnarray}
where we have assumed that ${\bf B}_{ext}$ is spatially constant (i.e., $\delta{\bf B}_{ext} = 0$). The terms $\hterm$ and $\zeta$, defined in Appendix \ref{sec:deltaBrho}, are higher order terms related to the gradient in the smoothing length which are necessary for strict conservation of energy -- hence we retain them here -- though they are generally expected to be negligible in practise.
Taking the perturbation (\ref{eq:deltarhoB}) with respect to $\delta x^{i}_{a}$ and using tensor notation gives
\begin{eqnarray}
\frac{\delta (\rho_{b} B_{b}^{j})}{\delta x^{i}_{a}} & = & 
\frac{\epsilon_{jkl}}{\Omega_{b}}\sum_{c} m_{c} (A^{b}_{k} - A^{c}_{k}) \left[(\delta_{ba} - \delta_{ca})\pder{}{x^{i}_{b}}\right] \pder{W_{bc} (h_{b})}{x^{l}_{b}} 
+ \frac{\epsilon_{jkl}}{\Omega_{b}} \sum_{c} m_{c} \left(\frac{\delta A^{b}_{k}}{\delta x^{i}_{a}} - \frac{\delta A^{c}_{k}}{\delta x^{i}_{a}}\right) \pder{W_{bc} (h_{b})}{x^{l}_{b}} + B_{ext}^{j}\frac{\delta \rho_{b}}{\delta x^{i}_{a}} \nonumber \\
& + & \left[ \htensor^{j}_{b} + \frac{B^{j}_{b,int}}{\Omega_{b}} \zeta_{b} \right] \frac{\delta\rho_{b}}{\delta x^{i}_{a}} + \frac{B^{j}_{b,int}\rho_{b}}{\Omega_{b}}\pder{h_{b}}{\rho_{b}} \sum_{c} m_{c} \left[(\delta_{ba} - \delta_{ca})\pder{}{x^{i}_{b}}\right] \pder{W_{bc}(h_{b})}{h_{b}}. \label{eq:deltarhoBtensor}
\end{eqnarray}

The perturbation to the vector potential $\delta{\bf A}$ can be expressed as a function of $\delta{\bf x}$, from (\ref{eq:dAdtSPH}), by
\begin{equation}
\delta A^{b}_{k} = \frac{A^{b}_{m}}{\Omega_{b}\rho_{b}} \sum_{d} m_{d} (\delta x^{m}_{b} - \delta x^{m}_{d}) \pder{W_{bd} (h_{b})}{x^{k}_{b}} + \epsilon_{kmn} \delta x^{m}_{b} B_{ext,b}^{n}.
\label{eq:deltaA}
\end{equation}

 At this point it is worth briefly pausing to examine the consequences of (\ref{eq:deltarhoB}) for the equations of motion. In particular if we consider (\ref{eq:deltarhoB}) and (\ref{eq:deltaA}) together, ignoring the last two terms relating to smoothing length gradients, then one may observe that there are essentially three separate terms that will contribute to the force. We will refer to these as the `2D', `external' and `3D' force terms. The 2D term arises from the first term in equation (\ref{eq:deltarhoB}) and follows only from the SPH formulation of the curl used to construct $\bf B$ from $\bf A$ (i.e., Equation \ref{eq:curlA}). We refer to this as the 2D term because it is the \emph{only} term which is present for purely two dimensional simulations where $\delta{\bf A} = 0$ (i.e., there is no external field). The `external' term is present in the case of external fields and arises from the combination of the $\delta {\bf x}\times {\bf B}_{ext}$ term in (\ref{eq:deltaA}) and the second term in Equation (\ref{eq:deltarhoB}), plus the third term from Equation (\ref{eq:deltarhoB}). Finally the 3D term arises from the combination of our choice both of gauge and SPH formulation of the induction equation (i.e., the first term of equation \ref{eq:deltaA}) and the curl operator via the second term in (\ref{eq:deltarhoB}).
 
  Whist the derivation of the 2D force is simply a matter of substituting the first term in (\ref{eq:deltarhoBtensor}) into (\ref{eq:deltaLexpanded}) and simplifying, the external and 3D force terms are more complicated since they involve first substituting the terms in (\ref{eq:deltaA}) into the second term of (\ref{eq:deltarhoBtensor}) and in turn into (\ref{eq:deltaLexpanded}).

\subsubsection{2.5D/${\bf B}_{ext}$ component}
  For the `external' force term, we can substitute the second term in (\ref{eq:deltaA}) into the second term of (\ref{eq:deltarhoB}) and add the third term from (\ref{eq:deltarhoB}) to obtain,
\begin{equation}
\left.\delta (\rho_{b} B_{b}^{j})\right._{ext} =  \frac{\epsilon_{jkl} \epsilon_{kmn}}{\Omega_{b}} \sum_{c} m_{c} \left( \delta x^{m}_{b} B^{n}_{ext,b} - \delta x^{m}_{c} B^{n}_{ext,c}  \right)\pder{W_{bc} (h_{b})}{x^{l}_{b}} + B_{ext}^{j}\delta\rho_{b}.
\end{equation}
Using the standard identity for the Levi-Civita tensor $\epsilon_{jkl} \epsilon_{kmn} = \delta_{jn}\delta_{lm} - \delta_{jm}\delta_{ln}$ we have
\begin{equation}
\left.\delta (\rho_{b} B_{b}^{j})\right._{ext} = 
   \frac{1}{\Omega_{b}} \sum_{c} m_{c} \left( \delta x^{l}_{b} B^{j}_{ext,b} - \delta x^{l}_{c} B^{j}_{ext,c}  \right)\pder{W_{bc} (h_{b})}{x^{l}_{b}}
- \frac{1}{\Omega_{b}} \sum_{c} m_{c} \left( \delta x^{j}_{b} B^{l}_{ext,b} - \delta x^{j}_{c} B^{l}_{ext,c}  \right)\pder{W_{bc} (h_{b})}{x^{l}_{b}}
+ B_{ext}^{j}\delta\rho_{b}.
\end{equation}
The astute reader will note that Galilean invariance in the perturbation (and thus momentum conservation in the equations of motion) \emph{only} follows if we make the simplifying assumption that the external magnetic field is spatially constant (and therefore independent of the particle positions). This is in fact physical since an external field with spatial gradients can impart momentum to the fluid and the equations in that case would not be expected to show Galilean invariance. In this paper we will deal only with spatially constant external fields, for which the above simplifies to (in vector form, where we have also substituted for $\delta \rho_{b}$ using \ref{eq:deltarhovec})
\begin{equation}
\left.\delta (\rho_{b} {\bf B}_{b})\right._{ext} =   
\frac{2{\bf B}_{ext}}{\Omega_{b}} \sum_{c} m_{c} \left( \delta {\bf x}_{b} - \delta {\bf x}_{c} \right)\cdot \nabla_{b} W_{bc} (h_{b})
- \frac{1}{\Omega_{b}} \sum_{c} m_{c} \left( \delta {\bf x}_{b} - \delta {\bf x}_{c} \right){\bf B}_{ext}\cdot \nabla_{b} W_{bc} (h_{b}).
\end{equation}
Taking the above perturbation with respect to $\delta x^{i}_{a}$, the resultant term in (\ref{eq:deltarhoBtensor}) is given by
\begin{equation}
\left[\frac{\delta (\rho_{b} B_{b}^{j})}{\delta x^{i}_{a}}\right]_{ext} = \frac{2B^{j}_{ext}}{\Omega_{b}} \sum_{c} m_{c} \pder{W_{bc} (h_{b})}{x^{i}_{b}} \left( \delta_{ba} - \delta_{ca} \right)
- \frac{\delta^{j}_{i} B^{l}_{ext}}{\Omega_{b}} \sum_{c} m_{c} \pder{W_{bc} (h_{b})}{x^{l}_{b}} \left( \delta_{ba} - \delta_{ca} \right).
\label{eq:externalterm}
\end{equation}

\subsubsection{3D component}
 The 3D force contribution is given by substituting the first term in (\ref{eq:deltaA}), taken with respect to $\delta x^{i}_{a}$, into the second term of (\ref{eq:deltarhoBtensor}), giving
\begin{eqnarray}
\left[\frac{\delta (\rho_{b} B_{b}^{j})}{\delta x^{i}_{a}}\right]_{3D} &= &
\frac{\epsilon_{jkl}}{\Omega_{b}} \sum_{c} m_{c} \frac{A^{b}_{i}}{\Omega_{b}\rho_{b}} \left[\sum_{d} m_{d}\pder{W_{bd}(h_{b})}{x^{k}_{b}} (\delta_{ba} - \delta_{da})\right] \pder{W_{bc} (h_{b})}{x^{l}_{b}}  \nonumber \\
& - & \frac{\epsilon_{jkl}}{\Omega_{b}} \sum_{c} m_{c} \frac{A^{c}_{i}}{\Omega_{c}\rho_{c}}\left[\sum_{d} m_{d}\pder{W_{cd}(h_{c})}{x^{k}_{c}} (\delta_{ca} - \delta_{da})\right] \pder{W_{bc} (h_{b})}{x^{l}_{b}}.
\label{eq:3Dterm}
\end{eqnarray}
The fact that the $\delta{\bf x}$ is so deeply nested in the perturbation of ${\bf B}$ in the 3D case (i.e., via a summation for the curl of $\bf A$ [equation \ref{eq:deltarhoB}], and via a second summation for $\delta{\bf A}$ [equation \ref{eq:deltaA}]), as we will see below, leads to a force term which is somewhat complicated to calculate. Nevertheless, it is a force term which preserves the basic symmetries that we asked for, namely momentum and energy conservation in the SPMHD equations. For example, using the naive or `standard' gauge choice (Equation \ref{eq:standardgauge}) involves one fewer summations since the $\delta {\bf x}$ is not nested inside a derivative in the perturbation to $\bf A$. However, the perturbation is not Galilean invariant and can be straightforwardly shown to lead to a force that does not conserve momentum.

\subsubsection{Equations of motion}
\label{sec:equationsofmotion}
Putting the perturbations (\ref{eq:deltarho}) and (\ref{eq:deltarhoBtensor}) [the second term of which has been expanded into (\ref{eq:externalterm}) and (\ref{eq:3Dterm})] into (\ref{eq:deltaLexpanded}) we have 
\begin{eqnarray}
\int\left\{-m_a \frac{dv^{i}_a}{dt} \right. & - & \sum_{b} \frac{m_{b}}{\Omega_{b}}\left[\frac{P_{b}}{\rho_{b}^{2}} - \frac{3}{2\mu_0} \left(\frac{B_b}{\rho_b}\right)^2 + \frac{\xi_{b}}{\rho_{b}^{2}} \right]\sum_{c} m_{c}  \pder{W_{bc}(h_{b})}{x^{i}_{b}} (\delta_{ba} - \delta_{ca}) \nonumber \\
& - & \frac{1}{\mu_0} \sum_{b} \frac{m_{b}}{\Omega_{b}}\frac{B^{j}_b}{\rho_{b}^{2}} \epsilon_{jkl}\sum_{c} m_{c} (A^{b}_{k} - A^{c}_{k}) \pder{^{2} W_{bc} (h_{b})}{x^{i}_{b} \partial x^{l}_{b}}  (\delta_{ba} - \delta_{ca}) \nonumber \\
& - & \frac{1}{\mu_{0}} \sum_{b} \frac{m_{b}}{\Omega_{b}} \frac{B_{b}^{j} B_{int,b}^{j}}{\rho_{b}} \pder{h_{b}}{\rho_{b}} \sum_{c} m_{c}(\delta_{ba} - \delta_{ca}) \pder{^{2} W_{bc}(h_{b})}{x^{i}_{b}\partial h_{b}} \nonumber \\
& - & \frac{1}{\mu_0}\sum_{b} \frac{m_{b}}{\Omega_{b}} \frac{B^{j}_b}{\rho_{b}^{2}} \left[2\delta^{l}_{i} B^{j}_{ext}
- \delta^{j}_{i} B^{l}_{ext} \right] \sum_{c} m_{c} \pder{W_{bc} (h_{b})}{x^{l}_{b}} \left( \delta_{ba} - \delta_{ca} \right) \nonumber \\
& - & \frac{1}{\mu_0} \sum_{b} \frac{m_{b}}{\Omega_{b}}\frac{B^{j}_b}{\rho_{b}^{2}}  \epsilon_{jkl}  \sum_{c} m_{c} \frac{A^{b}_{i}}{\Omega_{b}\rho_{b}} \left[\sum_{d} m_{d}\pder{W_{bd}(h_{b})}{x^{k}_{b}} (\delta_{ba} - \delta_{da})\right] \pder{W_{bc} (h_{b})}{x^{l}_{b}}  \nonumber \\
& + & \left.    \frac{1}{\mu_0} \sum_{b} \frac{m_{b}}{\Omega_{b}}\frac{B^{j}_b}{\rho_{b}^{2}}\epsilon_{jkl} \sum_{c} m_{c} \frac{A^{c}_{i}}{\Omega_{c}\rho_{c}}\left[\sum_{d} m_{d}\pder{W_{cd}(h_{c})}{x^{k}_{c}} (\delta_{ca} - \delta_{da})\right] \pder{W_{bc} (h_{b})}{x^{l}_{b}}  \right\}\delta x^{i}_a \mathrm{dt} = 0,
\label{eq:varprinlong}
\end{eqnarray}
where we have collected the isotropic terms relating to smoothing length gradients into a single term by defining
\begin{equation}
\xi_{b} \equiv \frac{1}{\mu_{0}}\left[ B^{j}_{b}\htensor_{b}^{j} + B_{b}^{j}B_{int,b}^{j}\frac{\zeta_{b}}{\Omega_{b}} \right],
\label{eq:xi}
\end{equation}
where $\htensor^{j}$ and $\zeta$ are defined in Appendix \ref{sec:deltaBrho}. Since the perturbation $\delta x^{i}$ is arbitrary, upon simplification (\ref{eq:varprinlong}) implies that the principle of least action is satisfied by the equations of motion in the form
\begin{eqnarray}
\frac{dv^{i}_{a}}{dt} & = & -\sum_{b} m_{b} \left[ \frac{P_{a} - \frac{3}{2\mu_{0}} B_{a}^{2} + \xi_{a}}{\rho_{a}^{2}\Omega_{a}} \pder{W_{ab} (h_{a})}{x^{i}_{a}} + \frac{P_{b} - \frac{3}{2\mu_{0}} B_{b}^{2} + \xi_{b}}{\rho_{b}^{2}\Omega_{b}} \pder{W_{ab} (h_{b})}{x^{i}_{a}} \right] 
\hspace{1.2cm}\left.\phantom{\frac{1}{2}} \right\}\textrm{isotropic term}\nonumber \\
& - & \frac{1}{\mu_{0}} \sum_{b} m_{b}  \left[ \frac{B^{j}_{a}}{\Omega_{a} \rho_{a}^{2}}\epsilon_{jkl} (A_{k}^{a} - A_{k}^{b}) \pder{^{2} W_{ab}(h_{a})}{x^{i}_{a} \partial x^{l}_{a}} + \frac{B^{j}_{b}}{\Omega_{b} \rho_{b}^{2}} \epsilon_{jkl}(A_{k}^{a} - A_{k}^{b}) \pder{^{2} W_{ab}(h_{b})}{x^{i}_{a} \partial x^{l}_{a}}\right]
\hspace{0.2cm}\left.\phantom{\frac{1}{2}} \right\}\textrm{2D term}\nonumber \\
& - & \frac{1}{\mu_{0}} \sum_{b} m_{b} \left[\frac{B_{a}^{j} B_{int,a}^{j}}{\Omega_{a}\rho_{a}} \pder{h_{a}}{\rho_{a}}\pder{^{2} W_{ab}(h_{a})}{x^{i}_{a}\partial h_{a}} +  \frac{B_{b}^{j} B_{int,b}^{j}}{\Omega_{b}\rho_{b}} \pder{h_{b}}{\rho_{b}}\pder{^{2} W_{ab}(h_{b})}{x^{i}_{a}\partial h_{b}} \right]\hspace{1.8cm}\left.\phantom{\frac{1}{2}} \right\}\textrm{2D $\nabla h$ term} \nonumber \\
& - & \frac{1}{\mu_0}\left[2\delta^{l}_{i} B^{j}_{ext} - \delta^{j}_{i} B^{l}_{ext} \right]
\sum_{b} m_{b} \left[ \frac{B^{j}_a}{\Omega_{a} \rho_{a}^{2}}  \pder{W_{ab} (h_{a})}{x^{l}_{a}} + \frac{B^{j}_b}{\Omega_{b} \rho_{b}^{2}}  \pder{W_{ab} (h_{b})}{x^{l}_{a}} \right] 
\hspace{1.6cm}\left.\phantom{\frac{1}{2}} \right\}\textrm{2.5D/${\bf B}_{ext}$ term}\nonumber \\
& - & \sum_{b} m_{b} \left[ \frac{A^{a}_{i} }{\Omega_{a}\rho_{a}^{2}} J^{k}_{a} \pder{W_{ab} (h_{a})}{x^{k}_{a}} +  \frac{A^{b}_{i} }{\Omega_{b}\rho_{b}^{2}}J^{k}_{b} \pder{W_{ab} (h_{b})}{x^{k}_{a}} \right],
\hspace{4.0cm}\left.\phantom{\frac{1}{2}} \right\}\textrm{3D term}
\label{eq:force}
\end{eqnarray}
where the current $J^{k}$ is defined according to
\begin{equation}
J^{k}_{a} \equiv -\rho_{a} \frac{ \epsilon_{kjl}}{\mu_{0}} \sum_{b} m_{b}\left[ \frac{B^{j}_{a}}{\Omega_{a}\rho_{a}^{2}} \pder{W_{ab} (h_{a})}{x^{l}_{a}} + \frac{B^{j}_{b}}{\Omega_{b}\rho_{b}^{2}}  \pder{W_{ab} (h_{b})}{x^{l}_{a}}\right],
\label{eq:gradBx}
\end{equation}
noting that the swapping of indices in the permutation tensor $\epsilon_{jkl}\to \epsilon_{kjl}$ leads to a change of sign in the corresponding term in equation (\ref{eq:force}). Although the curl operator in the above definition (\ref{eq:gradBx}) is determined entirely by the variational principle, remarkably this is simply the standard SPH symmetric curl operator in the presence of a variable smoothing length (i.e., Equation \ref{eq:symmetriccurl}). The equations of motion can be expressed in a more compact notation by writing the isotropic, $2.5$D/${\bf B}_{ext}$ and 3D terms in terms of a stress tensor and the 2D and 2D $\nabla h$ terms in terms of differential operators, giving
\begin{equation}
\frac{dv^{i}_{a}}{dt} = \sum_{b} m_{b} \left[ \left(\frac{ \mathcal{S}^{ij}_{a}}{\rho_{a}^{2}\Omega_{a}}  + \frac{(A_{ab}\times B_{a})^{j}}{\mu_{0}\rho^{2}_{a}\Omega_{a}}\pder{}{x_{a}^{i}} +\psi_{a}\delta^{i}_{j}\pder{}{h_{a}} \right) \pder{W_{ab}(h_{a})}{x^{j}_{a}} +  \left(\frac{ \mathcal{S}^{ij}_{b}}{\rho_{b}^{2}\Omega_{b}} + \frac{(A_{ab}\times B_{b})^{j}}{\mu_{0}\rho^{2}_{b}\Omega_{b}}\pder{}{x_{a}^{i}} +\psi_{b}\delta^{i}_{j}\pder{}{h_{b}}\right) \pder{W_{ab}(h_{b})}{x^{j}_{a}} \right],
\label{eq:equationsofmotionstress}
\end{equation}
where ${\bf A}_{ab} \equiv {\bf A}_{a} - {\bf A}_{b}$ and we have defined
\begin{eqnarray}
\mathcal{S}^{ij} & \equiv & -P \delta^{ij} + \frac{1}{\mu_{0}}\left[B^{i}B^{j}_{ext} + \delta^{ij}\left(\frac32 B^{2}  - 2{\bf B}\cdot{\bf B}_{ext} - \xi \right)\right] -  A^{i} J^{j}, \\
\psi_{a} & \equiv  & -\frac{1}{\mu_{0}}\frac{{\bf B}\cdot{\bf B}_{int}}{\Omega_{a}\rho_{a}} \pder{h_{a}}{\rho_{a}}.
\end{eqnarray}
Note that because the 2D terms cannot be represented by a stress tensor, $\mathcal{S}^{ij}$ does \emph{not} represent the usual MHD stress tensor, since the Lorentz force in this case is composed of the divergence of the stress tensor plus the 2D terms.

For the case of a constant smoothing length, the equations of motion simplify to (in vector notation)
\begin{eqnarray}
\frac{d{\bf v}_{a}}{dt} & = & -\sum_{b} m_{b} \left( \frac{P_{a} - \frac{3}{2\mu_{0}} B_{a}^{2}}{\rho_{a}^{2}} + \frac{P_{b} - \frac{3}{2\mu_{0}} B_{b}^{2}}{\rho_{b}^{2}} \right) \nabla_{a} W_{ab} \nonumber \\
& - & \frac{1}{\mu_{0}} \sum_{b} m_{b} \left\{ \left( \frac{{\bf B}_{a}}{\rho_{a}^{2}} +  \frac{{\bf B}_{b}}{\rho_{b}^{2}} \right)\cdot \left[({\bf A}_{a} - {\bf A}_{b}) \times \nabla \right] \right\} \nabla_{a} W_{ab} \nonumber \\
& - & \frac{2}{\mu_{0}}\sum_{b} m_{b} \left( \frac{{\bf B}_a}{\rho_{a}^{2}} + \frac{{\bf B}_b}{\rho_{b}^{2}} \right)\cdot {\bf B}_{ext} \nabla_{a} W_{ab} + \frac{1}{\mu_{0}}\sum_{b} m_{b} \left( \frac{{\bf B}_a}{\rho_{a}^{2}} + \frac{{\bf B}_b}{\rho_{b}^{2}} \right){\bf B}_{ext} \cdot \nabla_{a} W_{ab} \nonumber \\ 
& - & \sum_{b} m_{b} \left[ \frac{{\bf A}_{a}}{\rho_{a}^{2}} {\bf J}_{a}\cdot\nabla_{a} W_{ab}  +   \frac{{\bf A}_{b}}{\rho_{b}^{2}}{\bf J}_{b}\cdot\nabla_{a} W_{ab} \right],
\label{eq:eomfixedh}
\end{eqnarray}
where
\begin{equation}
{\bf J}_{a} \equiv \frac{\left(\nabla\times{\bf B}\right)_{a}}{\mu_{0}} \equiv -\frac{\rho_{a}}{\mu_{0}} \sum_{b} m_{b} \left[ \frac{{\bf B}_{a}}{\rho_{a}^{2}} +  \frac{{\bf B}_{b}}{\rho_{b}^{2}} \right] \times \nabla_{a} W_{ab}.
\label{eq:curlBfixedh}
\end{equation}

 At this point it is worth stepping back to consider the SPH formulations encapsulated by the force terms in (\ref{eq:force}) or equivalently, (\ref{eq:equationsofmotionstress}) or (\ref{eq:eomfixedh}). The most fundamental question is whether or not the magnetic force terms in the equations of motion derived above indeed are a representation of the Lorentz force when translated to the continuum limit. Since the proof is somewhat involved, the details and a translation of each of the terms into continuum form are given in Appendix \ref{sec:translate}. Given that the equations of motion are indeed correct in the continuum limit, the following comments can be made about their numerical representation:
\begin{enumerate}
\item The isotropic term in (\ref{eq:force}) is similar in form to the hydrodynamic SPH force and the usual isotropic MHD force in SPH. However, in this case the magnetic term is subtracted from the hydrodynamic pressure which implies that this term may be unstable to the clumping (tensile) instability caused by negative pressures in the regime where $\frac{3}{2\mu_{0}}{B^{2}} > P$.
\item The 2D and 3D terms present a novel formulation for the anisotropic magnetic force in SPH (strictly these terms also contain part of the isotropic force term -- see Appendix \ref{sec:translate}). The 2D term vanishes for constant ${\bf A}$ and is perpendicular to ${\bf A}$ yet remarkably both the 2D and 3D terms conserve linear momentum exactly since they are antisymmetric in the particle index -- implying that $\sum_{a} m_{a} d{\bf v}_{a}/dt = 0$.
\item Calculation of the 2D term involves use of the second derivative of the SPH kernel which is problematic using the cubic spline because the second derivative has discontinuous gradients. However this can be resolved using smoother kernels.
\item For a purely external magnetic field the stress tensor becomes $\mathcal{S}^{ij} = -P\delta^{ij} + 1/\mu_{0} (B^{i}_{ext} B^{j}_{ext} - \delta^{ij} \frac12 B^{2}_{ext})$ which is identical to the usual conservative SPMHD force term \citep[as derived, e.g., by][]{pm04b}. This part of the force, whilst conservative, is unstable to the tensile instability when the external magnetic pressure exceeds the gas pressure. The solution proposed by \citet{pm05} for this case was to simply subtract the constant term $B^{i}_{ext} B^{j}_{ext}/\mu_{0}$ from the stress when a constant external magnetic field is imposed.
\item Calculation of the 3D term involves a triple summation over the particles --- first, to calculate the density and simultaneously the magnetic field according to (\ref{eq:curlA}); second to calculate ${\bf J}$ via (\ref{eq:gradBx}) or (\ref{eq:curlBfixedh}); and third, to calculate the force term. Use of this approach is therefore 1/3 more expensive than a standard SPMHD scheme.
\end{enumerate}

\subsubsection{Instability correction}
\label{sec:correct}
As implied by items i) and iv) in the above discussion, we find in \S\ref{sec:tests} that the consistent variational formulation of the vector potential equations of motion (\ref{eq:force}) are in practise highly unstable to the tensile instability known to plague conservative SPMHD formulations in the standard case where the induction equation for the magnetic field ${\bf B}$ is evolved \citep{pm85,pm04a,pm05}. Worse still, we find that the equations are unstable for lower values of the magnetic field strength compared to the gas pressure (i.e., higher plasma $\beta$) than in the standard case, consistent with our conjecture in item i), above. That is, we find that the 2D term in (\ref{eq:force}) does not provide any significant stabilising influence over the negative stress arising from the isotropic term.

 For this reason we consider below ways of correcting the force term such that the net stress is positive in order to stabilise the formulation against the tensile instability. An obvious approach is to simply revert to the hybrid scheme discussed in \S\ref{sec:hybrid}. Indeed in the end (\S\ref{sec:summary}) we conclude that this is the best approach to implementing the vector potential in SPH, though one is left with the same problems that the consistent formulation given in \S\ref{sec:variational} was constructed to solve, namely that energy is not conserved exactly and that there is no (Hamiltonian) constraint on the overall evolution of the system (see discussion in \S\ref{sec:hybrid}). To this argument it may be countered that the consistent approach is no better in this respect once the correction terms below are added.

 In this paper, we stabilise the vector potential formulation we use the method proposed by \citet{bot01}, namely adding the ``source term'' $-{\bf B}(\nabla\cdot{\bf B})/\rho$ to the (conservative) force, giving
\begin{equation}
\frac{dv^{i}}{dt} = \frac{dv^{i}}{dt} - B^{i} \sum_{b} m_{b} \left[ \frac{B^{j}_{a}}{\rho_{a}^{2}}\pder{W_{ab}(h_{a})}{x^{j}_{a}}  + \frac{B^{j}_{b}}{\rho_{b}^{2}}\pder{W_{ab}(h_{b})}{x^{j}_{a}}\right].
\label{eq:botcorrect}
\end{equation}
This method violates both the conservation of momentum and energy, but only to the extent that the numerical estimate of $\nabla\cdot{\bf B}$ according to the above is non-zero. Whilst the proof that this correction term indeed stabilises the standard SPMHD equations has been given by \citet{bot04}, it is unclear why it should also work for the vector potential formulation. Empirically, we find that subtracting the ${\bf B}(\nabla\cdot{\bf B})$ term can indeed stabilise the vector potential force, but only in a limited range of circumstances, the limitations of which are unclear. For example, the 1D shock tube problems (\S\ref{sec:shocktubes}) are stabilised effectively by this method and similarly the circularly polarised Alfv\'en wave in 2D (\S\ref{sec:alfven}). However the 2D Orszag-Tang Vortex (\S\ref{sec:orstang}) remains unstable even with the correction term added. Ideally a full stability analysis of the vector potential equations of motion (\ref{eq:force}) should be carried out, though such a task is well beyond the scope of this paper.

 A more severe alternative would be to use the original method of \citet{pm85}, namely to correct the stress by subtracting the maximum value over all the particles,
\begin{equation}
S^{ij} = S^{ij} - S_{max}.
\label{eq:smax}
\end{equation}
Whilst this method conserves momentum but not energy, the correction to the stress can become arbitrarily large. For the Orszag-Tang Vortex problem (\S\ref{sec:orstang}) we find that the stress correction required to stabilise the solution starts to produce unphysical features in the solution and is therefore an unacceptable alternative.

\section{Dissipative terms}
\label{sec:diss}
 In Papers~I and III the need to introduce dissipative terms in the magnetic field in order to account for discontinuities in ${\bf B}$ was discussed. For the magnetic field this means adding an artificial resistivity term. The key constraints on deriving such a term are that it should i) conserve total energy and ii) result in a positive definite increase in entropy -- or equivalently -- thermal energy. \citet{pm04a} used these constraints to derive appropriate artificial resistivity terms for the standard SPMHD approach (i.e., using ${\bf B}$ or ${\bf B}/\rho$ in the induction equation).

\subsection{Resistivity using the vector potential}
\label{sec:dissA}
  The formulation of resistivity in the vector potential formulation is considerably simplified since the derivatives of $\eta$ do not enter the evolution equation for the vector potential (c.f. equation~\ref{eq:dAdt}). The appropriate dissipative term in the vector potential evolution is therefore given by
\begin{equation}
\left(\frac{d{\bf A}_{a}}{dt}\right)_{diss} = - \eta_{a} {\bf J}_{a},
\label{eq:dAdtdiss}
\end{equation}
where the resistivity $\eta_{a}$ and an SPH expression for ${\bf J}$ remain to be defined.

 The constraint of total energy conservation is expressed by equation (\ref{eq:dEdt}). Since the magnetic dissipation (physically) does not enter the equations of motion nor the continuity equation, we are left with the requirement that
\begin{equation}
\sum_{a} m_{a} \left[ \left(\frac{du_{a}}{dt}\right)_{diss} + \frac{{\bf B}_{a}}{\mu_{0}\rho_{a}}\cdot \left(\frac{d{\bf B}_{a}}{dt} \right)_{diss}\right] = 0.
\end{equation}
 The reader should note that --- whilst we do not consider dissipative terms as part of the derivation of the equations of motion from the Lagrangian in \S\ref{sec:variational} --- the above would be equivalent to stating that the contribution to the Lagrangian from the perturbation to the magnetic evolution is exactly balanced by the contribution from the perturbation to internal energy (from an increase in entropy) in (\ref{eq:deltaL}), thus having no effect on the equations of motion.
Writing $d{\bf B}/dt$ in terms of $d{\bf A}/dt$ using (\ref{eq:appdBdt}) gives
\begin{equation}
\sum_{a} m_{a} \frac{du_{a}}{dt} = - \sum_{a} m_{a} \frac{{\bf B}_{a}}{\Omega_{a}\rho_{a}^{2}} \cdot \sum_{b} m_{b} \left(\frac{d{\bf A}_{a}}{dt} -  \frac{d{\bf A}_{b}}{dt}\right) \times \nabla_{a} W_{ab} (h_{a}),
\end{equation}
where the above expression is determined by our choice of SPH operator for $\nabla\times{\bf A}$ (see Appendix \ref{sec:deltaBrho}) and we have dropped the subscript $_{diss}$ from the time derivatives assuming we are referring to the magnetic dissipation terms only. Using (\ref{eq:dAdtdiss}) in the above gives
\begin{equation}
\sum_{a} m_{a} \frac{du_{a}}{dt} = \sum_{a} m_{a} \frac{{\bf B}_{a}}{\Omega_{a}\rho_{a}^{2}} \cdot \sum_{b} m_{b} \left(\eta_{a} {\bf J}_{a} -  \eta_{b} {\bf J}_{b}\right) \times \nabla_{a} W_{ab} (h_{a}).
\end{equation}
Adding half of this term to half of the same term with the indices $a$ and $b$ exchanged, using the antisymmetry of the kernel $\nabla_{b} W_{ba}(h_{b}) = -\nabla_{a} W_{ab} (h_{b})$ and the vector identity ${\bf B}\cdot (\eta {\bf J} \times\nabla W) = - \eta {\bf J} ({\bf B}\times \nabla W)$ gives
\begin{eqnarray}
\sum_{a} m_{a} \frac{du_{a}}{dt} & = & -\frac12 \sum_{a} m_{a} \sum_{b} m_{b} \left(\eta_{a} {\bf J}_{a}\right)  \cdot \left [ \frac{{\bf B}_{a}}{\Omega_{a}\rho_{a}^{2}} \times \nabla_{a} W_{ab} (h_{a}) + \frac{{\bf B}_{b}}{\Omega_{b}\rho_{b}^{2}} \times \nabla_{a} W_{ab} (h_{b})\right ] \nonumber \\
& & + \frac12 \sum_{a} m_{a} \sum_{b} m_{b} \left( \eta_{b} {\bf J}_{b}\right)  \cdot \left [ \frac{{\bf B}_{a}}{\Omega_{a}\rho_{a}^{2}} \times \nabla_{a} W_{ab} (h_{a}) + \frac{{\bf B}_{b}}{\Omega_{b}\rho_{b}^{2}} \times \nabla_{a} W_{ab} (h_{b})\right ].
\end{eqnarray}
Swapping summation indices in the second term and combining the two terms, we have,
\begin{equation}
\sum_{a} m_{a} \frac{du_{a}}{dt} = -\sum_{a} m_{a} \sum_{b} m_{b} \left(\eta_{a} {\bf J}_{a}\right)  \cdot \left [ \frac{{\bf B}_{a}}{\Omega_{a}\rho_{a}^{2}} \times \nabla_{a} W_{ab} (h_{a}) + \frac{{\bf B}_{b}}{\Omega_{b}\rho_{b}^{2}} \times \nabla_{a} W_{ab} (h_{b})\right ],
\end{equation}
giving the contribution to the thermal energy equation for particle $a$ as
\begin{equation}
\frac{du_{a}}{dt} = - \left(\eta_{a} {\bf J}_{a}\right)  \cdot \sum_{b} m_{b}\left [ \frac{{\bf B}_{a}}{\Omega_{a}\rho_{a}^{2}} \times \nabla_{a} W_{ab} (h_{a}) + \frac{{\bf B}_{b}}{\Omega_{b}\rho_{b}^{2}} \times \nabla_{a} W_{ab} (h_{b})\right ].
\label{eq:JdotJsym}
\end{equation}
If we define ${\bf J}$ in (\ref{eq:dAdtdiss}) using the symmetric curl (equation \ref{eq:symmetriccurl}), then we have simply
\begin{equation}
\left(\frac{du_{a}}{dt}\right)_{diss} =  \frac{\eta_{a} J^{2}_{a}}{\rho_{a}},
\label{eq:dudtdiss}
\end{equation}
which is exactly the continuum expression. Furthermore the dissipation is guaranteed to be positive definite so long as ${\bf J}$ in (\ref{eq:dAdtdiss}) (and in \ref{eq:dudtdiss}) is calculated using the symmetric curl operator.

 As an aside, it is interesting to note that, as in the equations of motion, in the dissipation we are required to use the symmetric curl operator for ${\bf J}$ to obtain energy conservation, where in this case our only choice of SPH formalism was to specify the operator used in ${\bf B} = \nabla\times{\bf A}$. The reason for the appearance of the symmetric curl is because the curl operators in (\ref{eq:curlA}) and (\ref{eq:symmetriccurl}) form a conjugate pair. This conjugacy in SPH operators has been noted earlier by \citet{cr99} in the context of divergence cleaning, though not with the variable smoothing length formulation.

 The disadvantage of the symmetric curl is that it can give a poor representation of ${\bf J}$ if the particles are disordered. In particular on the shock tube tests (\S\ref{sec:tests}) we find problems using the symmetric curl for the dissipation in combination with the consistent SPMHD equations of motion derived in \S\ref{sec:equationsofmotion}, though using exactly the same dissipation in combination with either (\ref{eq:morrisforce}) or (\ref{eq:botforce}) gives good results. A compromise approach is to use the usual curl operator (as in equation \ref{eq:BcurlA}) for ${\bf J}$ in (\ref{eq:dAdtdiss}) and (\ref{eq:dudtdiss}). Provided that $J^{2}$ is used in (\ref{eq:dudtdiss}) (rather than using \ref{eq:JdotJsym}) then the dissipation will be guaranteed to be positive definite, however energy conservation will only be approximate because the energy-conserving expression is given by (\ref{eq:JdotJsym}). Alternatively energy conservation can be enforced by using (\ref{eq:JdotJsym}), whereby positivity of the dissipation is only guaranteed so long as the ${\bf J}$ estimate used in (\ref{eq:dAdtdiss}) and the symmetric curl estimate of ${\bf J}$ given by the summation in (\ref{eq:JdotJsym}) have the same sign.


\subsection{Choosing $\eta$ corresponding to an artificial resistivity}
 The remaining issue is to formulate the resistivity parameter $\eta$ appropriately for an \emph{artificial} resistivity, that is where the resistivity acts only on the smallest scales in the calculation to diffuse discontinuities in the magnetic field. We propose the following:
\begin{equation}
\eta_{a} = \alpha_{B} v^{a}_{A} h_{a}, 
\label{eq:eta}
\end{equation}
where $\alpha_{B}$ is a dimensionless factor of order unity, $v^{a}_{A}$ is the Alfv\'en speed and $h$ is the particle's smoothing length. An alternative which is second order in the smoothing length would be to use
\begin{equation}
\eta_{a} = \alpha_{B} \sqrt{\frac{\mu_{0} J^{2}}{\rho_{a}}} h^{2}_{a}, 
\label{eq:etaJ}
\end{equation}
where ${\bf J}$ is the current density. This gives a resistive diffusion that is nonlinear in the smoothing length and responds only to large gradients in ${\bf B}$. In general we find (\S\ref{sec:tests}) that (\ref{eq:etaJ}) gives insufficient dissipation at discontinuities, though perhaps some combination of (\ref{eq:eta}) and (\ref{eq:etaJ}) could be a reasonable compromise.

\subsection{Comparison with previous formulations}
\label{sec:disscompare}
  The above formulation for the vector potential differs considerably from that required in the usual SPMHD approach \citep{pm04a} since $\eta$ cannot be defined using the pairwise signal velocity between particles and their mutual separation. The derivation given above also shows that the naive dissipative terms previously formulated by \citet{rp07} and \citet{pb07} for the vector/Euler potentials, involving the pairwise signal velocity, have no guarantee of positive definite contributions to the thermal energy. The dissipation term used by \citet{rp07} was given by
\begin{equation}
\left(\frac{d{\bf A}_{a}}{dt}\right)_{diss} = \sum_{b} \frac{m_{b}}{\bar{\rho}_{ab}} \alpha_{B} v_{sig} \left( {\bf A}_{a} - {\bf A}_{b} \right)\overline{F_{ab}},
\label{eq:dAdtdisswrong}
\end{equation}
where $\overline{F_{ab}} = \frac12 \left[F_{ab}(h_{a}) + F_{ab}(h_{b})\right]$ is the scalar part of the kernel gradient term (see Appendix~\ref{sec:kernelderivs}). Following the analysis given above, it can be shown that the appropriate term to be added to the thermal energy equation in order to conserve total energy is given by
\begin{equation}
\left(\frac{du_{a}}{dt}\right)_{diss} =  -\frac12 \sum_{b} m_{b} \frac{\alpha_{B} v_{sig}}{\bar{\rho}_{ab}} \left({\bf J}_{a} - {\bf J}_{b}\right)\cdot ({\bf A}_{a} - {\bf A}_{b}) \overline{F_{ab}}.
\end{equation}
The dot product between ${\bf J}_{ab}$ and ${\bf A}_{ab}$ has no guarantee of being positive definite, and indeed in numerical tests on the shock tube problems discussed in \S\ref{sec:tests} we find that negative thermal energies can result. Instead \citet{rp07} and \citet{pb07} add the term used in the usual SPH formulation by \citet{pm05}, i.e.,
\begin{equation}
\left(\frac{du_{a}}{dt}\right)_{diss} =  -\frac12 \sum_{b} m_{b} \frac{\alpha_{B} v_{sig}}{\mu_{0}\bar{\rho}_{ab}} \left({\bf B}_{a} - {\bf B}_{b}\right)^{2}\overline{F_{ab}},
\end{equation}
which is positive definite, but when combined with (\ref{eq:dAdtdisswrong}) has no guarantee of conserving energy. Thus, this approach should be discarded in favour of the correct formulation given by equations (\ref{eq:dAdtdiss}) and (\ref{eq:dudtdiss}). Note that (\ref{eq:dAdtdiss}) and (\ref{eq:dudtdiss}) should be used regardless of whether the equations of motion derived in \S\ref{sec:equationsofmotion} are implemented. That is, even using an alternative representation for the force terms such as (\ref{eq:morrisforce}), the dissipation for the vector potential should still be implemented using equations (\ref{eq:dAdtdiss}) and (\ref{eq:dudtdiss}), the expressions for ${\bf J}$ in which are determined entirely by the choice of curl operator in ${\bf B} = \nabla\times{\bf A}$ (see \S\ref{sec:BcurlA}). There are similar implications for the dissipative terms used with the Euler potentials, though these will be discussed elsewhere.

\section{Numerical tests}
\label{sec:tests}
 We have implemented the vector potential formulation of SPMHD into the $N$-dimensional SPH code (hereafter, {\sc ndspmhd}) that we have previously used to test the standard SPMHD formulation in \citet{price04} and Papers~I--III. The code evolves the SPMHD equations using a standard leapfrog predictor-corrector scheme, where the vector potential ${\bf A}$ is evolved alongside the velocity field. The timestep is controlled globally as described in \citet{pm05}, where unless specified we use a Courant factor of $C_{cour} = 0.30$.

\begin{figure*}
\begin{center}
\includegraphics[angle=270,width=0.9\textwidth]{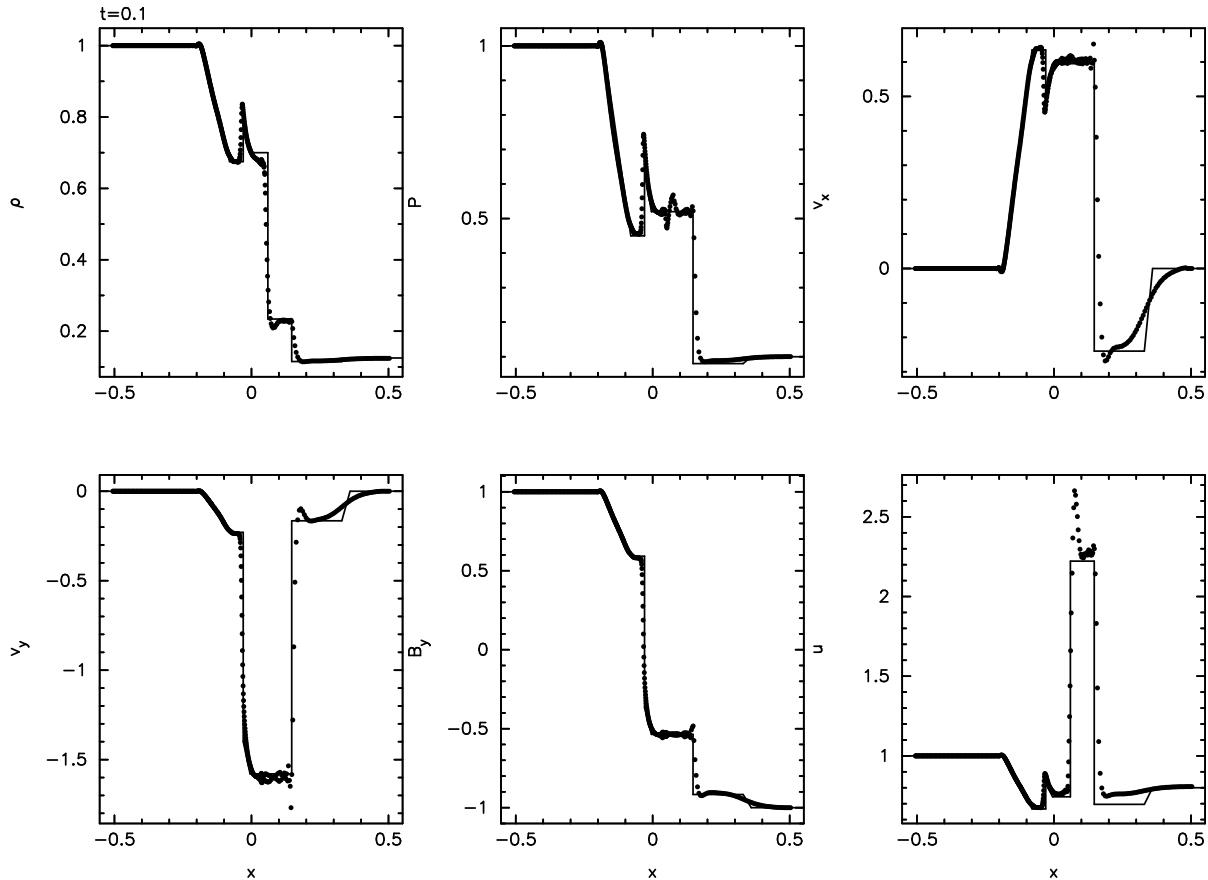}
\caption{Results of the \citet{bw88} shock tube problem using the consistent vector potential formulation (\ref{eq:force}), corrected for stability using (\ref{eq:botcorrect}), using the quintic kernel instead of the cubic spline and replacing the symmetric curl operator required in the MHD dissipation terms with a more accurate but non-conservative estimate. Artificial viscosity, conductivity and resistivity have been applied, the first of these using a switch, the second using the formulation of \citet{price08} and the third as in \S\ref{sec:diss} using $\alpha_{B} = 0.75$.}
\label{fig:briowuvecp}
\end{center}
\end{figure*}

\begin{figure*}
\begin{center}
\includegraphics[angle=270,width=0.9\textwidth]{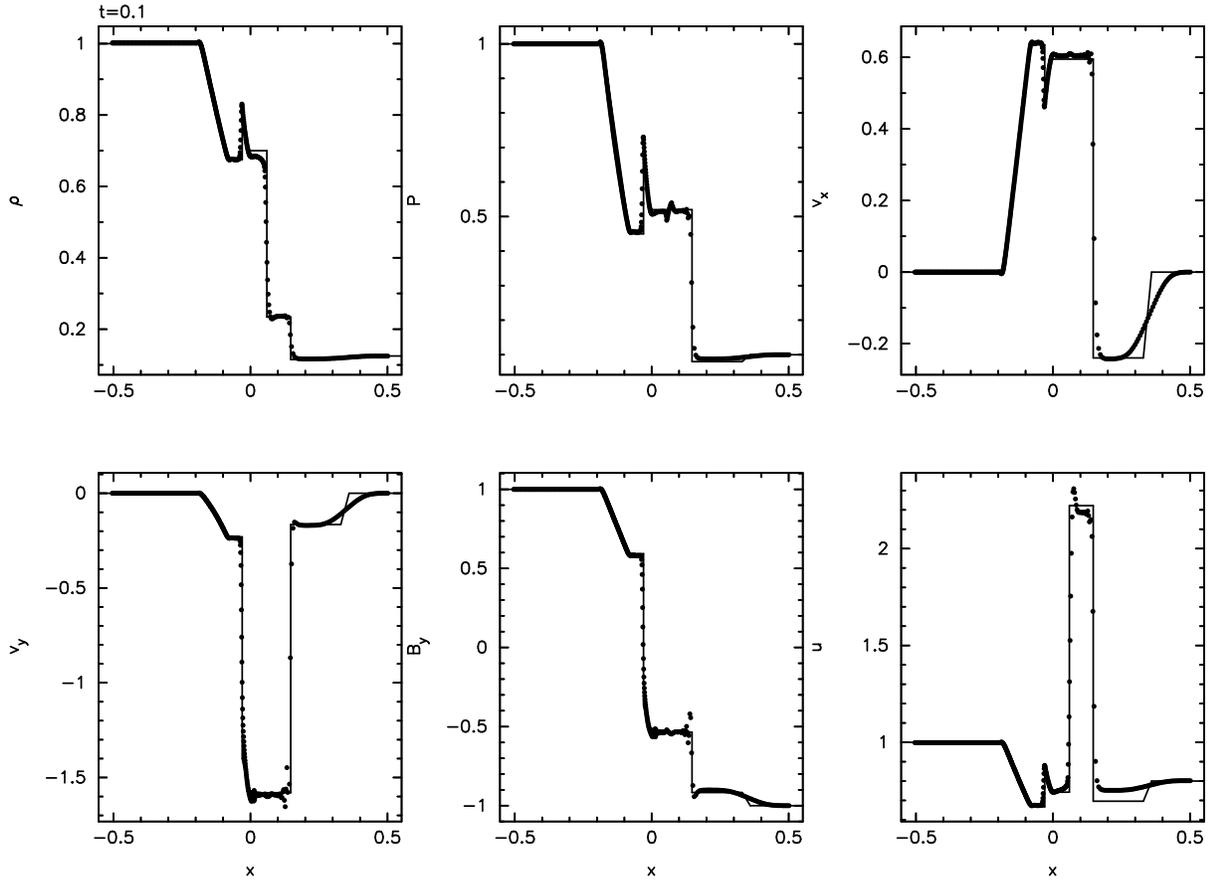}
\caption{Results of the \citet{bw88} shock tube probelm using a hybrid but non-conservative approach. In this case we have been able to use the usual cubic spline kernel and the correct curl formulation to ensure that the MHD dissipation is both positive definite and energy-conserving. Artificial viscosity, conductivity and resistivity have been applied as in Figure~\ref{fig:briowuvecp}, though with a slightly lower resistivity parameter ($\alpha_{B} = 0.5$).}
\label{fig:briowuhybrid}
\end{center}
\end{figure*}
 
  Use of the vector potential compared to the standard variable-smoothing length SPMHD scheme \citep{pm05} involves the following changes to the code:
\begin{enumerate}
\item During the iterated loop over the particles to calculate both density, smoothing length and $\Omega$ from (\ref{eq:omega}), calculate ${\bf B}$ using (\ref{eq:curlA}) and the smoothing length gradient terms $\hterm$ and $\zeta$ using (\ref{eq:hsum}) and (\ref{eq:zeta}) respectively. Note that $\hterm$ and $\zeta$ can be combined with ${\bf B}$ at the end of the summation loop to construct $\xi$ according to (\ref{eq:xi}) and thus stored only as a single scalar variable.
\item In three dimensions an extra loop over the particles is required to calculate the current ${\bf J}$ using the symmetric curl (equation~\ref{eq:gradBx}).
\item The force is calculated in the main loop according to (\ref{eq:force}), alongside which the time derivative of the vector potential is evaluated using (\ref{eq:dAdtSPH}), with dissipation according to (\ref{eq:dAdtdiss}) and (\ref{eq:dudtdiss}).
\item Where external fields are used, boundary (ghost) particles require that the boundary value of the vector potential evolves according to the second term in (\ref{eq:dAdtSPH}).
\end{enumerate}

 In the following sections we examine the performance of the consistent formulation of the vector potential on a range of test problems that are commonly used to test (both grid based and SPH) MHD codes. The tests are identical to those described in Papers~I--III apart from the 3D version of the Orzsag-Tang Vortex which has also been considered by \citet{ds08}. Our first aim is to compare the energy-conserving or ``consistent'' formulation derived in \S\ref{sec:variational} both to the standard SPMHD scheme \citep{pm05} but also to the more naive hybrid approach (\S\ref{sec:hybrid}) used in previous papers \citep{pb07,rp07}, where in this paper we use (\ref{eq:botforce}) though results are similar with the Morris force. For this purpose, one and two dimensional problems suffice since the equations of motion (\ref{eq:force}) already differ from the hybrid approach using (\ref{eq:botforce}) or (\ref{eq:morrisforce}) in one dimension. Shock tube problems are particularly valuable for assessing the role of dissipative terms, for which the formulation used in this paper also differs from those used previously (see \S\ref{sec:disscompare}), as well as the effect of the ``2.5D'' terms in (\ref{eq:force}) relating to an external magnetic field. The Orszag-Tang Vortex in 2D, though the most complicated in terms of dynamics, is the simplest test problem in terms of implementation, since the only non-zero terms present from (\ref{eq:force}) are the isotropic and 2D terms.

 Our second aim is to examine the accuracy of the vector potential in general, using either the energy-conserving (``consistent'') or hybrid approach.  Testing of the vector potential evolution equation (\ref{eq:dAdtSPH}) as distinct from the Euler Potentials (\S\ref{sec:intro}) requires a three dimensional test problem, for which we consider a 3D version of the Orszag-Tang Vortex. Since we find that this problem cannot be effectively stabilised using the consistent formulation, we consider it only with the hybrid approach.

\begin{figure*}
\begin{center}
\includegraphics[angle=270,width=\textwidth]{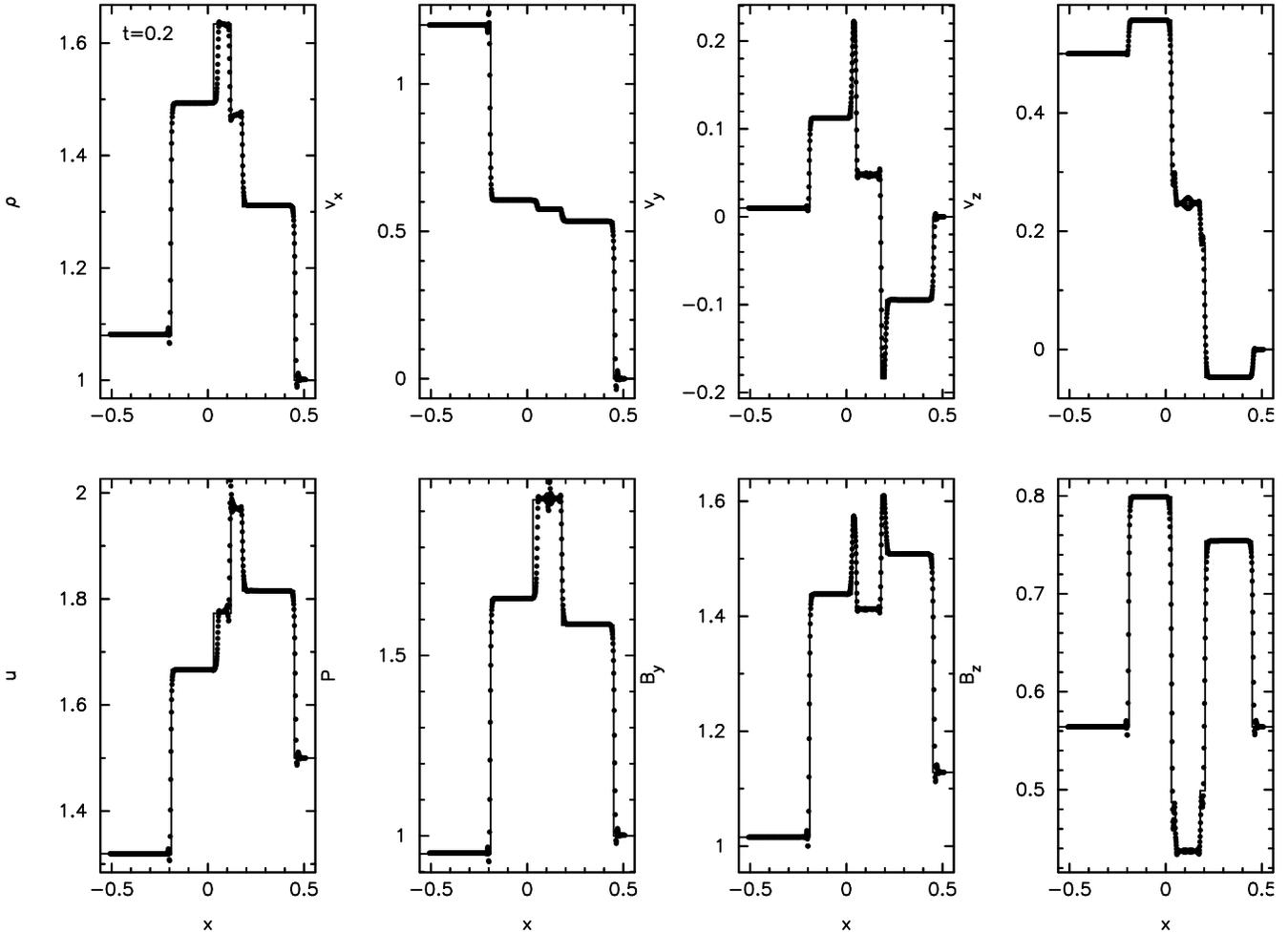}
\caption{Results of the adiabatic MHD shock tube problem showing the formation of seven distinct discontinuities related to the propagation of MHD waves. The solution has been computed using the consistent vector potential force, corrected for stability with (\ref{eq:botcorrect}), using the standard cubic spline kernel and an artificial resistivity parameter $\alpha_{B} = 0.15$.}
\label{fig:mshk3}
\end{center}
\end{figure*}

\begin{figure*}
\begin{center}
\includegraphics[angle=270,width=0.9\textwidth]{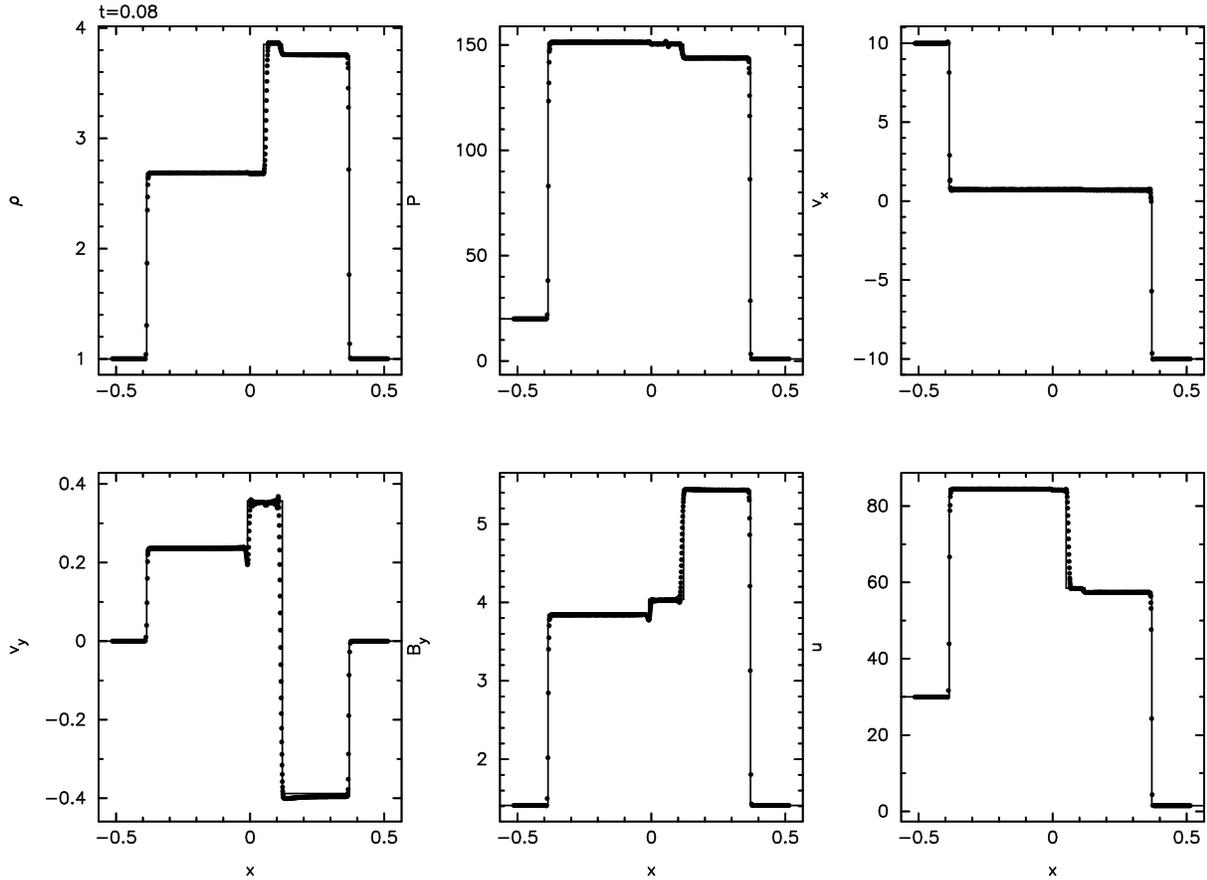}
\caption{Results using the consistent vector potential formulation of the MHD shock tube problem with a strong compression and weaker shearing discontinuities. Here we have added the correction to the force term for stability, the standard cubic spline kernel and an more accurate but non-conservative curl in the MHD dissipation terms which are applied using a switch. The results may be compared with the exact solution given by the solid line.}
\label{fig:mshk7}
\end{center}
\end{figure*}

\subsection{1.5D shock tube problems}
\label{sec:shocktubes}
 The shock tube described by \citet{bw88} is perhaps the most widely used test problem for MHD codes. We set up the problem exactly as described in \citet{pm04b}, with no smoothing of the initial conditions and using approximately 800 equal mass particles in the domain $x = [-0.5, 0.5]$. Conditions to the left of the shock are given by $(\rho,P,v_x,v_y,B_y) = [1,1,0,0,1]$ and to the right by $(\rho,P,v_x,v_y,B_y)=[0.125,0.1,0,0,-1]$ with $B_x = 0.75$ and $\gamma=2.0$. For the vector potential this means that we set ${\bf B}_{ext} = [B_{x},0,0]$ and $A_{z} = -B_{y} x$ initially.
 
  The results using the consistent formulation of the vector potential force (equation \ref{eq:force}) are shown in Figure~\ref{fig:briowuvecp} and may be compared to the exact solution given by the solid line. There are numerous issues. To produce a reasonable solution \emph{at all} on this problem, we have had to:
\begin{itemize}
\item Stabilise the force term using (\ref{eq:botcorrect}),
\item use the quintic kernel (\ref{eq:quintic}) instead of the cubic spline (\ref{eq:cubic}), and
\item use a more accurate ${\bf J}$ estimate in the dissipation term (\ref{eq:dAdtdiss}) rather than the symmetric curl.
\end{itemize}
 The solution shown in Figure~\ref{fig:briowuvecp} also requires a relatively high dissipation parameter ($\alpha_{B} = 0.75$) that is \emph{not} applied using a switch. The fact that we have not used the symmetric curl means that the dissipation term does not conserve energy exactly --- we have instead ensured that the contribution to the entropy is positive definite (see \S\ref{sec:dissA}). We also checked whether or not simply subtracting the constant external component from the stress tensor as discussed in point iv) of \S\ref{sec:equationsofmotion} would stabilise the result instead of (\ref{eq:botcorrect}) but found this not to be the case.
  
 The results using the hybrid approach (i.e., using (\ref{eq:botforce}) for the force) are shown in Figure~\ref{fig:briowuhybrid} and were obtained with far fewer tweaks --- that is, using the cubic spline, $\alpha_{B} = 0.5$ and the dissipation applied using the symmetric curl for ${\bf J}$ as required for energy conservation. The remaining issues with over-smoothing of the fast rarefaction are similarly present in the standard SPMHD formulation \citep[see][]{pm04b} and are mainly due to the fact that the resolution is very low in this region due to the density jump. The oscillations around the slow magnetosonic shock can be calmed further by increasing the resistivity parameter, though at the expense of further smoothing the fast rarefaction. The results in this case are similar to those presented by \citet{rp07} which is to be expected since the only difference is in the correct formulation of the dissipative terms used here. The results in Figures~\ref{fig:briowuvecp} and \ref{fig:briowuhybrid} are an improvement over those presented in \citet{pm04a} and \citet{pm04b} although this is mainly because of a better understanding of how to apply dissipation terms developed in \citet{pm05} rather than being an intrinsic improvement due to the use of the vector potential.


 As the \citet{bw88} problem is more difficult in SPH because of the density contrast, we consider two further shock tube problems in this paper, both of which have also been used to test the standard SPMHD formulation \citep{price04,pm04a}.
 
  The second shock tube illustrates the formation of seven discontinuities in the same problem (Figure~\ref{fig:mshk3}). The Riemann problem is set up with initial conditions to the left ($x < 0$) of the shock given by $(\rho,P,v_x,v_y,v_z,B_y,B_z) = [1.08,0.95,1.2,0.01,0.5,3.6/(4\pi)^{1/2},2/(4\pi)^{1/2}]$, whilst to the right ($x > 0$) $(\rho,P,v_x,v_y,v_z,B_y,B_z)=[1,1,0,0,0,4/(4\pi)^{1/2},2/(4\pi)^{1/2}]$ with $B_x = 2/(4\pi)^{1/2}$ and $\gamma=5/3$. Using the vector potential we set ${\bf B}_{ext} = [B_{x}, 0, B_{z}]$ and $A_{z} = -B_{y} x$ initially. Since the velocity in the x-direction is non-zero at the boundary, we continually inject particles into the left half of the domain with the appropriate left state properties. The resolution therefore varies from an initial 700 particles to 875 particles at $t=0.2$. The results are shown in Figure \ref{fig:mshk3} at time $t=0.2$, using the consistent vector potential force formulation. As with the Brio-Wu problem, we find that it is necessary to correct the force term according to (\ref{eq:botcorrect}) in order to obtain stability, though in this case the solution is obtained satisfactorily using the cubic spline kernel, the symmetric curl in the dissipation and a low resistivity parameter $\alpha_{B} = 0.15$. Figure~\ref{fig:mshk3} may be compared with Figure~4 in \citet{pm04a}, though as in the Brio-Wu problem the differences compared to that paper are more due to the improvements made in \citet{pm05} rather than anything specific to the vector potential. Results with the hybrid formulation (\S\ref{sec:hybrid}) are similar. 

 The final shock tube problem we consider has initial conditions to the left of the shock given by $(\rho,P,v_x, v_y,B_y) =
[1,20,10,0,5/(4\pi)^{1/2}]$ and to the right by $(\rho,P,v_x,v_y,B_y) = [1,1,-10,0,5/(4\pi)^{1/2}]$ with $B_x = 5.0/(4\pi)^{1/2}$  and $\gamma = 5/3$. The vector potential is set up using ${\bf B}_{ext} = [B_{x}, 0, 0]$ and $A_{z} = -B_{y} x$ initially. The results computed using the consistent vector potential formulation are shown at $t=0.08$ in Figure~\ref{fig:mshk7} and may be compared with the exact solution taken from \citet{dw94} given by the solid line. We have used $800$ particles in the domain. Figure~\ref{fig:mshk7} may be compared with Figures 4.18 and 4.19 of \citet{price04} for the standard SPMHD formulation. For this problem, apart from applying the stability correction (\ref{eq:botcorrect}) we have nonetheless used the cubic spline kernel, though with the consistent formulation satisfactory results could not be obtained using the symmetric curl in the dissipation and thus we have reverted to the more accurate (asymmetric) ${\bf J}$ estimate, forgoing exact energy conservation. Here we have applied both artificial viscosity and resistivity using the switches discussed in \citet{pm05}, and artificial conductivity as described in \citet{price08} with $\alpha_{u} = 1$.

\subsection{2.5D Circularly Polarised Alfv\'en wave}
\label{sec:alfven}
 The circularly polarised Alfv\'en wave is an exact, non-linear solution of the MHD equations. It is particularly useful as a test problem as it allows one to compute the evolution of a non-linear wave of arbitrary amplitude indefinitely, since the wave does not compress the gas and therefore does not steepen into a shock. The parameters for the test problem used here are identical to those described by \citet{toth00} for Eulerian codes and the setup for SPH is identical to that described in Paper~III for the standard SPMHD scheme except that here we set up the magnetic field in terms of the vector potential. 
 
 Whilst the reader is referred to \citet{pm05} for further details, we briefly recap the setup parameters: The wave is setup in a two dimensional domain with a unit wavelength along the direction of propagation (ie. in this case along the line at an angle of $30^\circ$ with respect to the x-axis). The initial conditions are $\rho = 1$, $P = 0.1$, $v_\parallel = 0$, $B_\parallel = 1$, $v_\perp = B_\perp = 0.1\sin{(2\pi r_\parallel)}$ and $v_z = B_z = 0.1 \cos{(2\pi r_\parallel)}$ with $\gamma = 5/3$ (where $r_\parallel = x \cos{\theta} + y\sin{\theta}$). The $x-$ and $y-$ components of the magnetic field are therefore given by $B_x = B_\parallel\cos{\theta} - B_\perp \sin{\theta} $ and $B_y = B_\parallel \sin{\theta} + B_\perp \cos{\theta}$ (and similarly for the velocity and vector potential components). Conversely, $B_\parallel = B_y \sin{\theta} + B_x \cos{\theta}$ and $B_\perp = B_y \cos{\theta} - B_x \sin{\theta}$.
 
  For the vector potential we set up two components, $A_{\perp} = 0.1[ \sin{(2\pi r_{\Vert})}]/2\pi$ and $A_{z} = 0.1[ \cos{(2\pi r_{\Vert})}]/ 2\pi$, together with the $B_\parallel$ component added as an external field. As the external field is the dominant component of the magnetic field, this is mainly a test of the 2D and 2.5D/$B_{ext}$ terms in (\ref{eq:force}) together with the evolution of the magnetic field according to (\ref{eq:dAdtSPH}). Furthermore, since the magnetic pressure exceeds the gas pressure, this test is unstable to the SPH tensile instability in the standard formulation (Paper~I). Using our consistent vector potential formulation (equation \ref{eq:force}) we expect the force to be unstable in the regime where $P > 3/2 B^{2}$ (see note i. in \S\ref{sec:equationsofmotion}). In order to assess whether or not this was true in practice we computed a series of tests adjusting the value of the gas pressure (since the test is non-compressive and the wave travels at the Alfv\'en speed, adjusting the gas pressure should have no effect on the results). Indeed we find that, in the absence of instability corrections the test is indeed unstable to the tensile instability and that this occurs for all simulations where $P \lesssim 3$.


\begin{figure*}
\begin{center}
\includegraphics[width=0.49\textwidth]{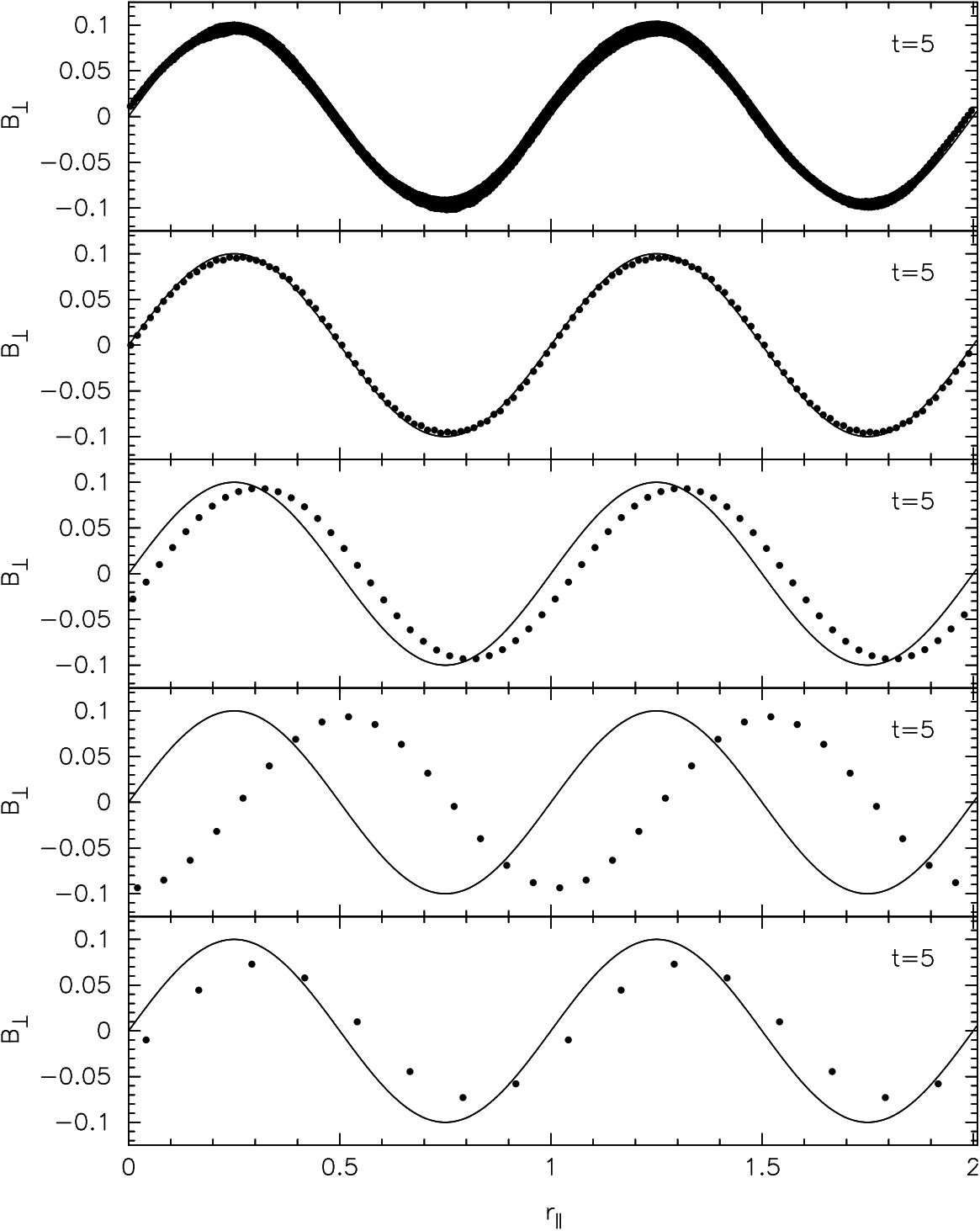}
\hspace{0.005\textwidth}
\includegraphics[width=0.49\textwidth]{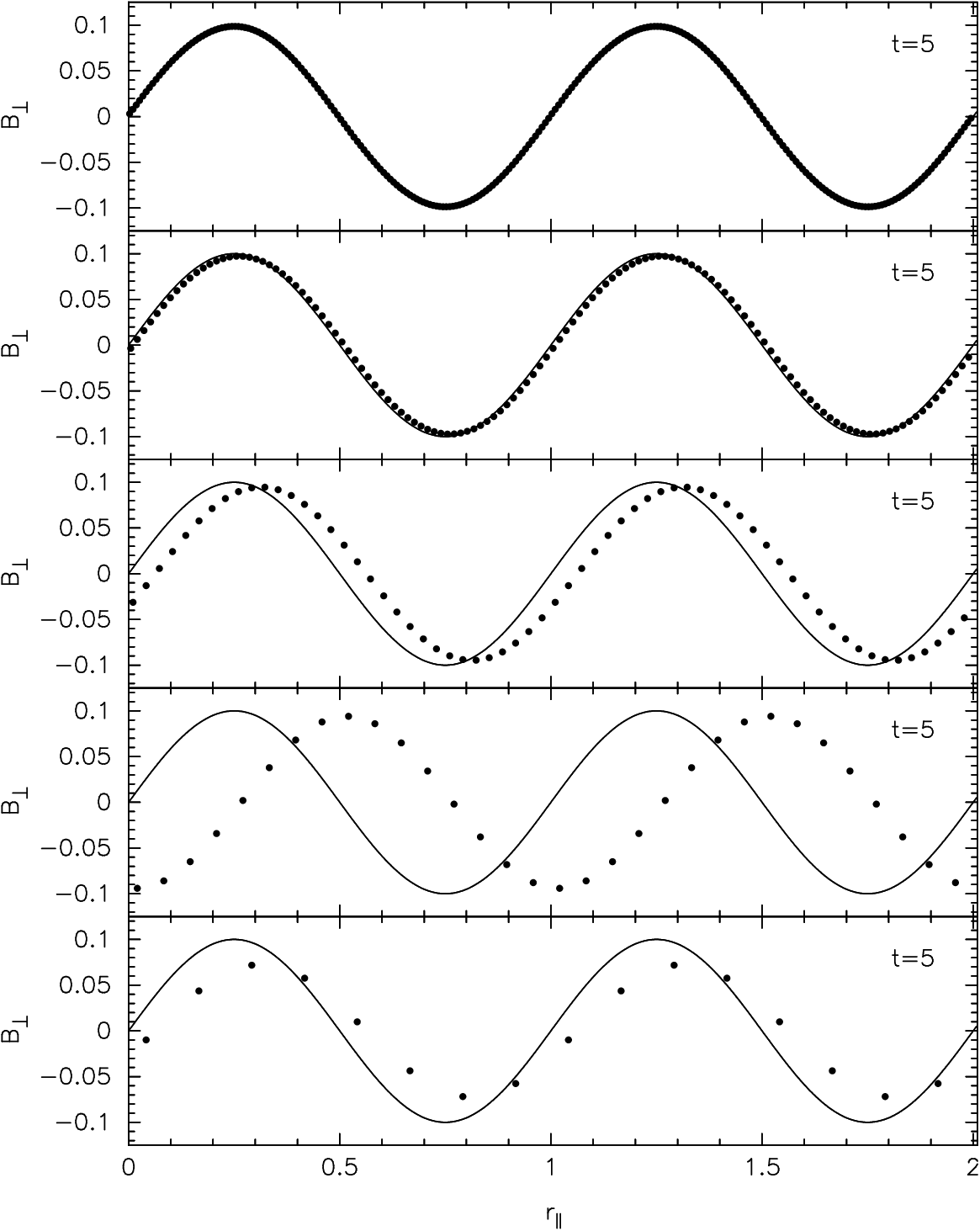}
\caption{Results of the circularly polarized Alfv\'en wave test at $t=5$ (corresponding to 5
wave periods). The plots show the perpendicular component of the magnetic field vector
$B_\perp = B_y \cos{\theta} - B_x \sin{\theta}$ for all of the particles,
projected against a vector parallel to the direction of wave propagation
$r_\parallel = x \cos{\theta} + y\sin{\theta}$ (where $\theta = 30^\circ$ in this
case). The SPMHD results are shown at five different resolutions which are, from
bottom to top, $8\times 16$, $16\times 32$, $32\times 64$, $64\times 128$ and
$128\times 256$. The exact solution is given by the solid line. The left panel shows the results using the conservative vector potential formulation with the constant external stress subtracted whilst the right panel shows results using a `hybrid approach' -- that is, evolving the vector potential but with a standard SPMHD force term. Artificial viscosity was applied using a switch, though no artificial resistivity. The results are indistinguishable except that the highest resolution calculation in the left panel starts to show noise in the particle distribution at late times related to the excitation of compressible modes, most likely related to the poor accuracy in the kernel second derivatives.}
\label{fig:malfvenresults}
\end{center}
\end{figure*}

\subsection{2D Orszag-Tang vortex}
\label{sec:orstang}
 The compressible version of the \citet{ot79} vortex has also been widely used as a test problem for both grid based and SPH MHD codes \citep[e.g.][]{rea95,dw98,ld00,toth00,pm05,rp07}.The setup we use, identical to that in \citet{rea95} and \citet{ld00} consists of an initially uniform density, periodic $1 \times 1$ box given an initial velocity perturbation ${\bf v} = v_0[-\sin{(2\pi y)},\sin{(2\pi x)}]$ where $v_0 = 1$. The magnetic field is given a doubly periodic geometry ${\bf B} = B_0[-\sin{(2\pi y)},\sin{(4\pi x)}]$ where $B_0 = 1/\sqrt{4\pi}$, which for the vector potential is achieved by setting $A_{z} = -B_{0}/\pi \left[  \frac12 \cos{(2\pi y)} +  \frac14 \cos{(4\pi x)}\right]$ initially. No external fields are present for this problem. The flow has an initial average Mach number of unity, a ratio of magnetic to thermal pressure of $\beta = 10/3$ and ratio of specific heats $\gamma = 5/3$. The initial gas state is therefore $P = 5/3 B_0^2 = 5/(12\pi)$ and $\rho = \gamma P/v_0 = 25/(36\pi)$. In this paper we have used $128^{2}$ particles, initially placed on a cubic lattice, which is sufficient to demonstrate energy conservation and the problems with numerical stability faced by the energy-conserving formulation.

 We have considered the problem in its original 2D form and also in a 3D geometry following \citet{ds08}. As discussed above the Orszag-Tang Vortex in 2D is the simplest test problem for the vector potential in terms of implementation, since it is purely two dimensional --- in the magnetic field (i.e., no external fields) as well as the spatial coordinates. As there is no inflow at the boundaries it is also a good test for checking energy conservation in the code.
 
\begin{figure*}
\begin{center}
\includegraphics[angle=270,width=\textwidth]{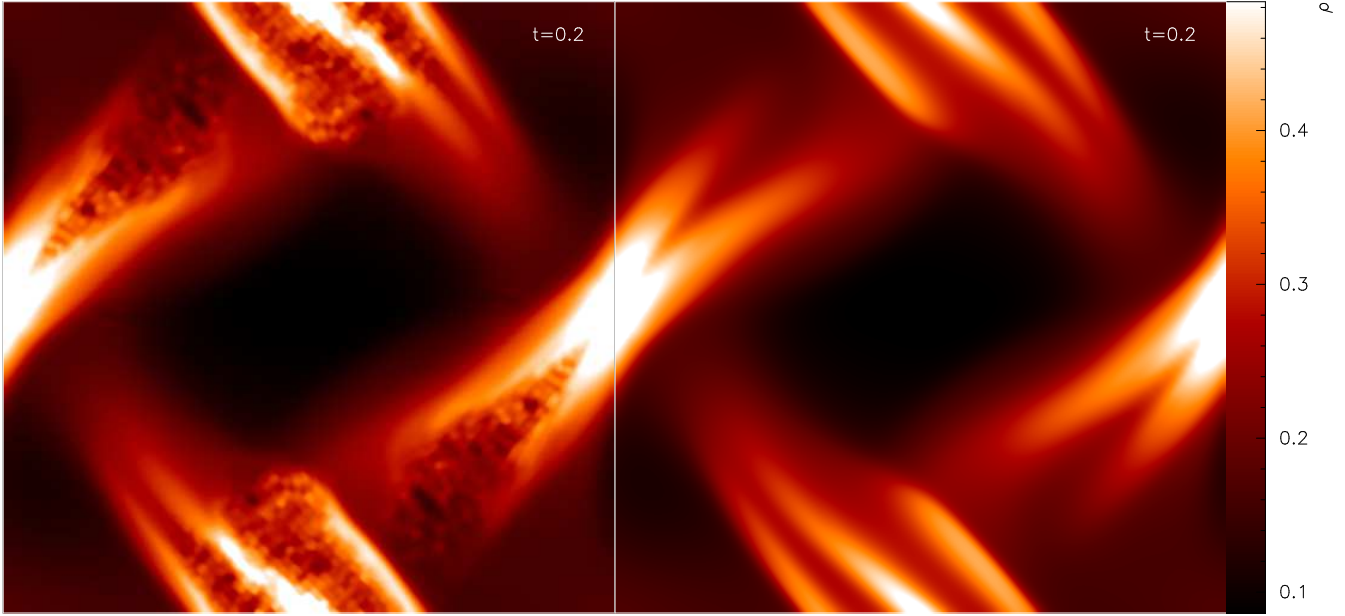}
\caption{Density in the 2D Orszag-Tang vortex problem at $t=0.2$ in a $128^{2}$ particle calculation, showing development of the numerical tensile instability in the energy-conserving formulation of the vector potential (left), compared to a hybrid approach using a stable but non-conservative force formulation (right). The instability develops in the low density regions of the flow where the magnetic pressure exceeds gas pressure. Whilst the standard SPMHD formulation evolving ${\bf B}$ suffers from similar problems in the $\beta < 1$ regime when a conservative force is used, the onset of the instability for the conservative vector potential formulation occurs at lower magnetic field strengths ($\beta \lesssim 3$).}
\label{fig:orstangbad}
\end{center}
\end{figure*}

 We find that the calculation using the energy-conserving vector potential formulation becomes unstable to the tensile instability at relatively early times ($t\gtrsim0.1$), manifested by the usual symptom of particles clumping along the field lines in the low density, high magnetic pressure regions. The development of the instability is clear by $t=0.2$ as shown in the density field in the left panel of Figure~\ref{fig:orstangbad}. Worse still, adding the correction term (\ref{eq:botcorrect}) does \emph{not} stabilise the instability for this problem. In fact the only fail-safe method we have found of removing the instability in this case is to subtract the maximum value from the stress tensor as in (\ref{eq:smax})--- and we find that doing so in this case produces unphysical artefacts in the solution. Thus we are unable to produce a stable and accurate solution to the Orszag-Tang Vortex using the consistent vector potential formulation.
 
\begin{figure}
\includegraphics[angle=270,width=0.45\textwidth]{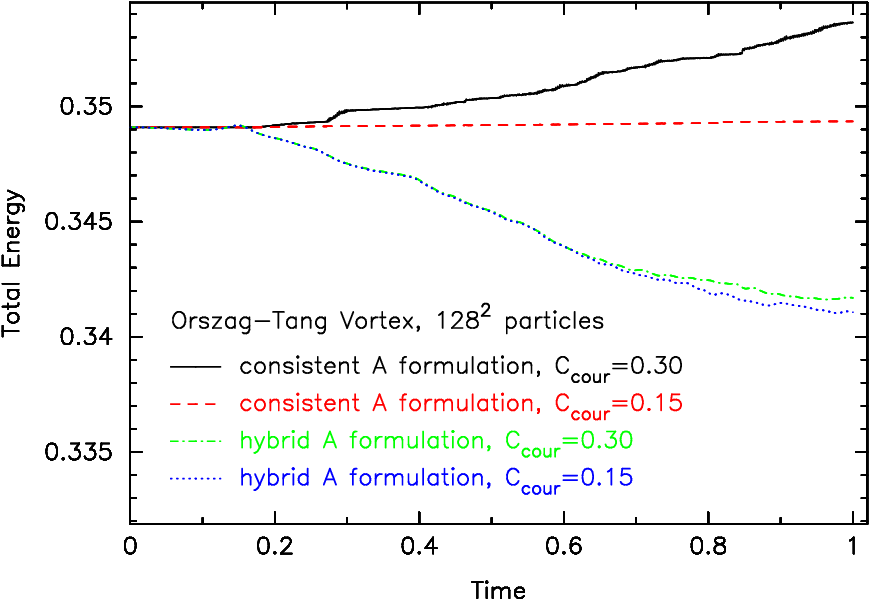}\hspace{0.02\textwidth}
\includegraphics[angle=270,width=0.45\textwidth]{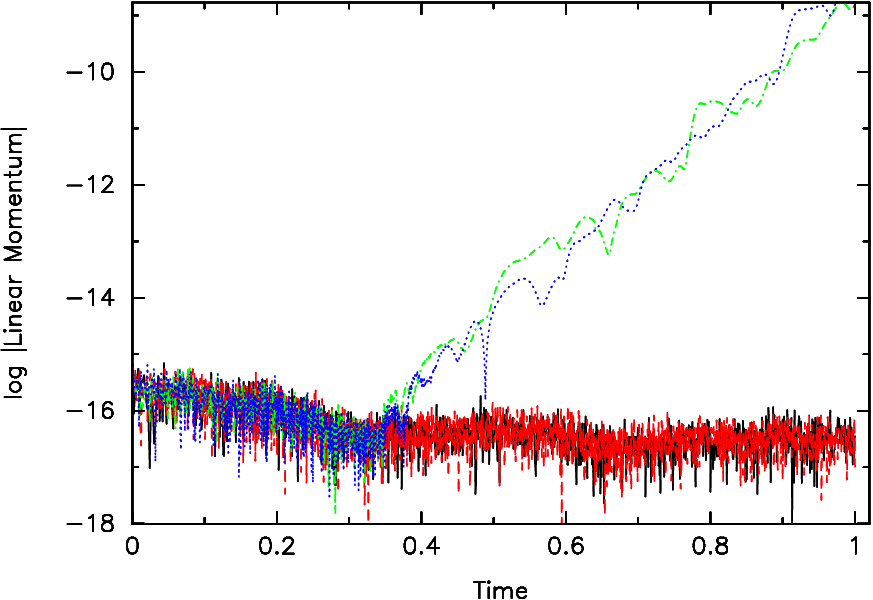}
\caption{Energy conservation in the 2D Orzsag-Tang problem, comparing $128^{2}$ particle calculations using the consistent formulation (solid/black, dashed/red lines) to the hybrid approach (dot-dashed/green, dotted/blue lines), for two different settings for the Courant factor in the timestepping ($C_{cour}$ as indicated). The corresponding density field at $t=0.2$ is shown in Figure~\ref{fig:orstangbad}. Despite the strong numerical instabilities present in the solution with the consistent formulation (Figure~\ref{fig:orstangbad}), energy is conserved exactly to the accuracy of the timestepping scheme and linear momentum (right panel) is conserved to machine precision. Total energy and momentum are not conserved regardless of the timestep with the hybrid approach because of the non-conservative source term in (\ref{eq:botforce}). However, the solution does not suffer from numerical instabilities (Figure~\ref{fig:orstangbad}, right panel; \citealt{rp07}).}
\label{fig:orstangen}
\end{figure}

 Using the hybrid approach the solution has no such difficulties (right panel), identical to the results presented in \citet{rp07}. What is remarkable though is that despite the severe numerical instability present with the consistent formulation, energy is in fact much better conserved than using the hybrid scheme, as demonstrated in Figure~\ref{fig:orstangen}. This figure shows the evolution of the total energy (left) and magnitude of the linear momentum (right) for the two calculations shown in Figure~\ref{fig:orstangbad} and also for two calculations using a smaller Courant factor. For the consistent formulation, it can be seen from Figure~\ref{fig:orstangbad} that energy is conserved exactly, i.e., to timestepping accuracy --- meaning that the energy conservation can be improved to arbitrary precision by decreasing the Courant factor. By contrast, decreasing the timestep further in the hybrid case produces essentially no change in the energy conservation, indicating that the non-conservation derives from the SPH scheme itself. Indeed we find that this is a rigorous test of the implementation of the energy-conserving formulation in the code, since even by neglecting the high-order terms relating to the smoothing length gradients discussed in Appendix~\ref{sec:deltaBrho} we find that the total energy rises rapidly to unphysical values once the instability sets in.

 In fact, a very good solution to the 2D Orszag-Tang Vortex problem can be obtained using the hybrid approach, as has already been demonstrated by \citet{rp07}. We do not feel that it is necessary to repeat those results here, referring the reader to \citet{rp07} for details. Instead, in this paper we skip directly to the three dimensional version of the problem since we are interested in differences between using the vector potential compared with the Euler potentials, in particular the effect of solving (\ref{eq:dAdt}) instead of (\ref{eq:eulerevol}).

\subsection{3D Orszag-Tang vortex}
\label{sec:orszagtang3D}
\begin{figure*}
\includegraphics[angle=270,width=0.45\textwidth]{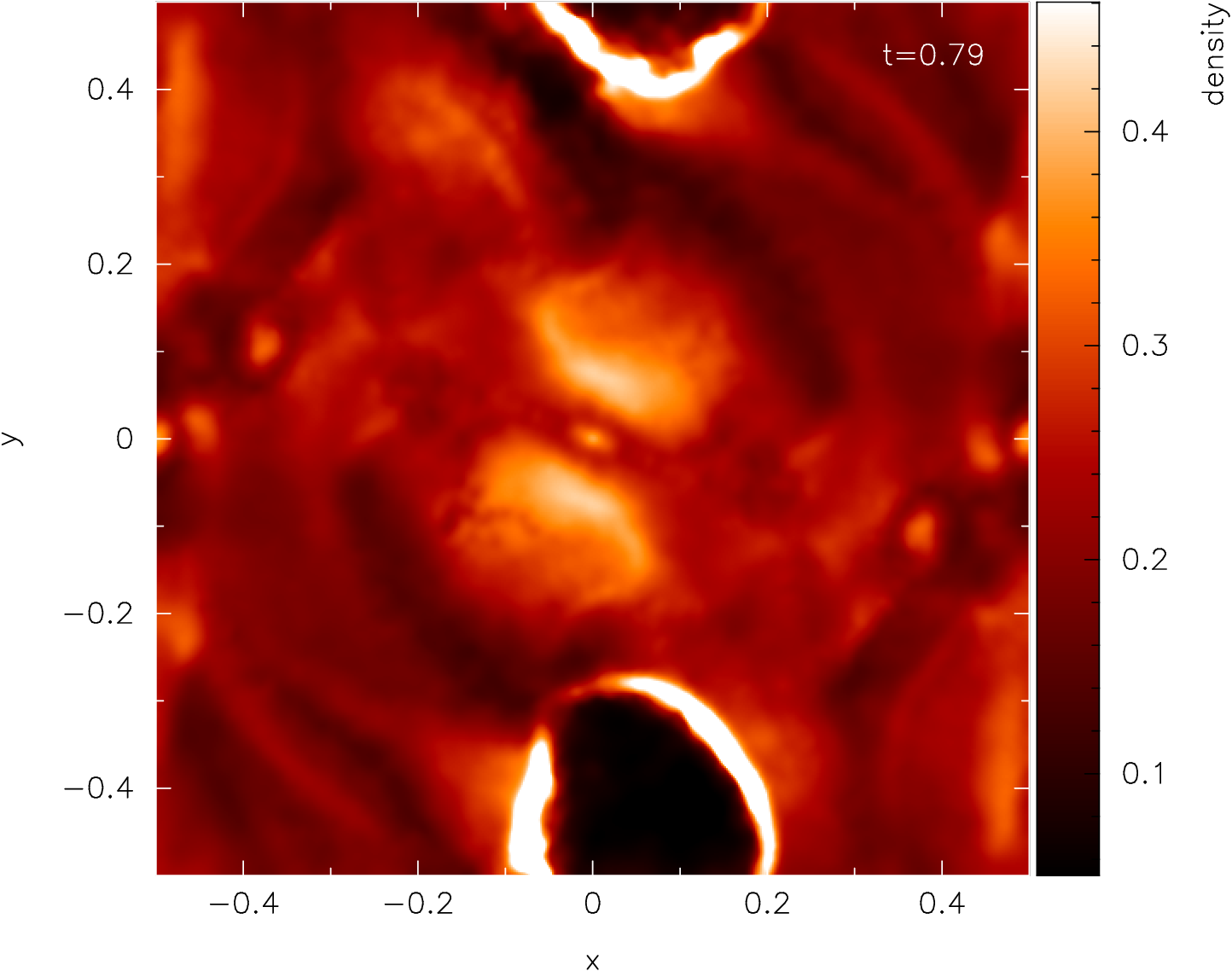}
\hspace{0.01\textwidth}
\includegraphics[angle=270,width=0.37\textwidth]{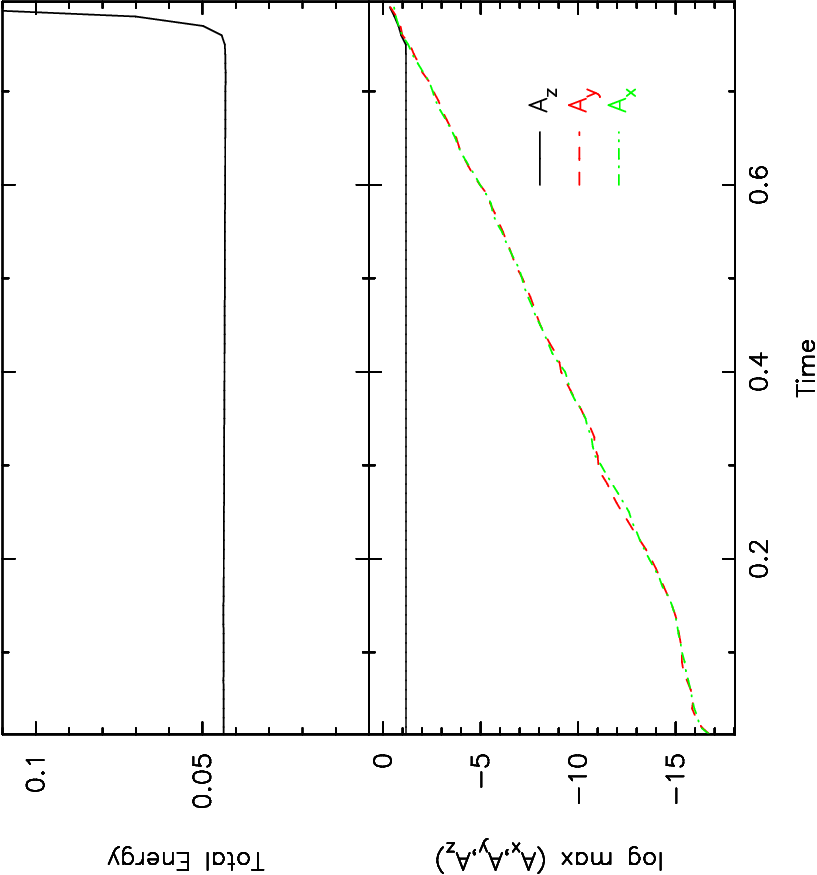}

\includegraphics[angle=270,width=0.45\textwidth]{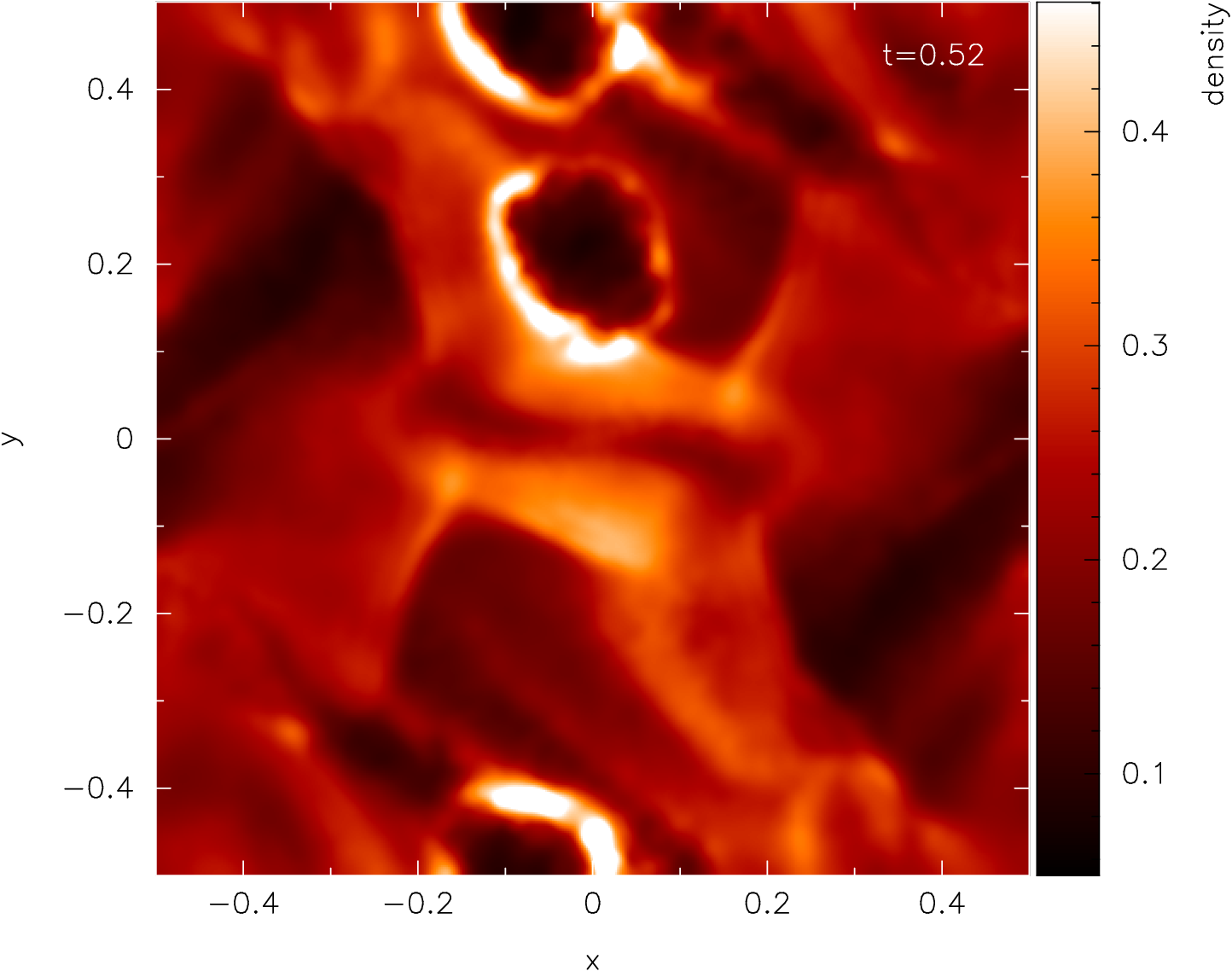}
\hspace{0.01\textwidth}
\includegraphics[angle=270,width=0.37\textwidth]{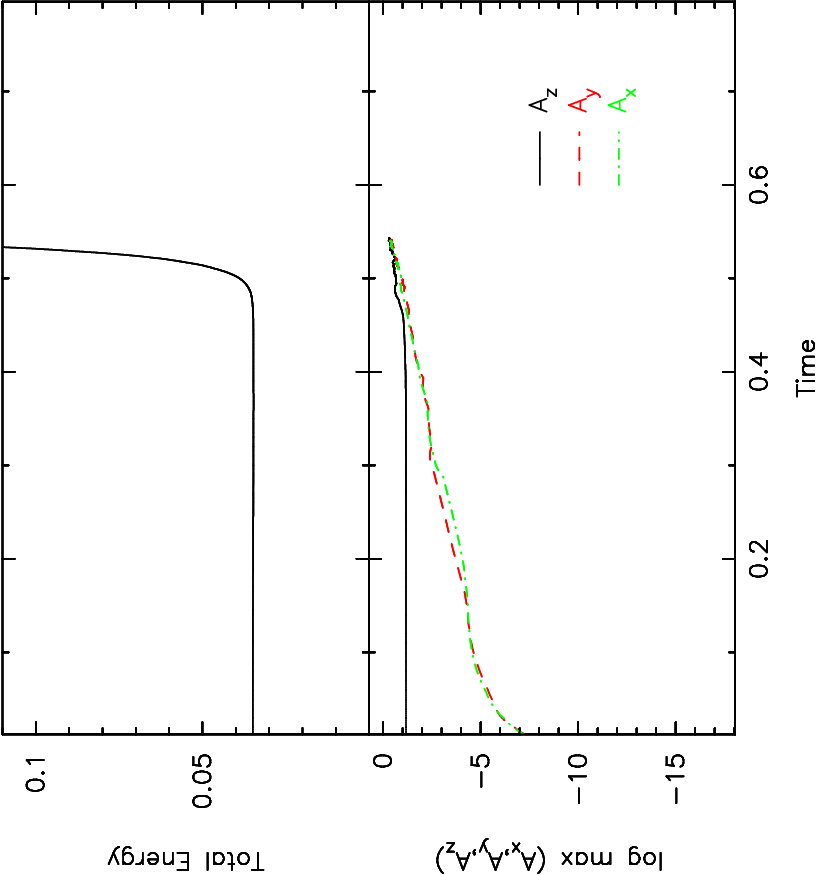}

\includegraphics[angle=270,width=0.45\textwidth]{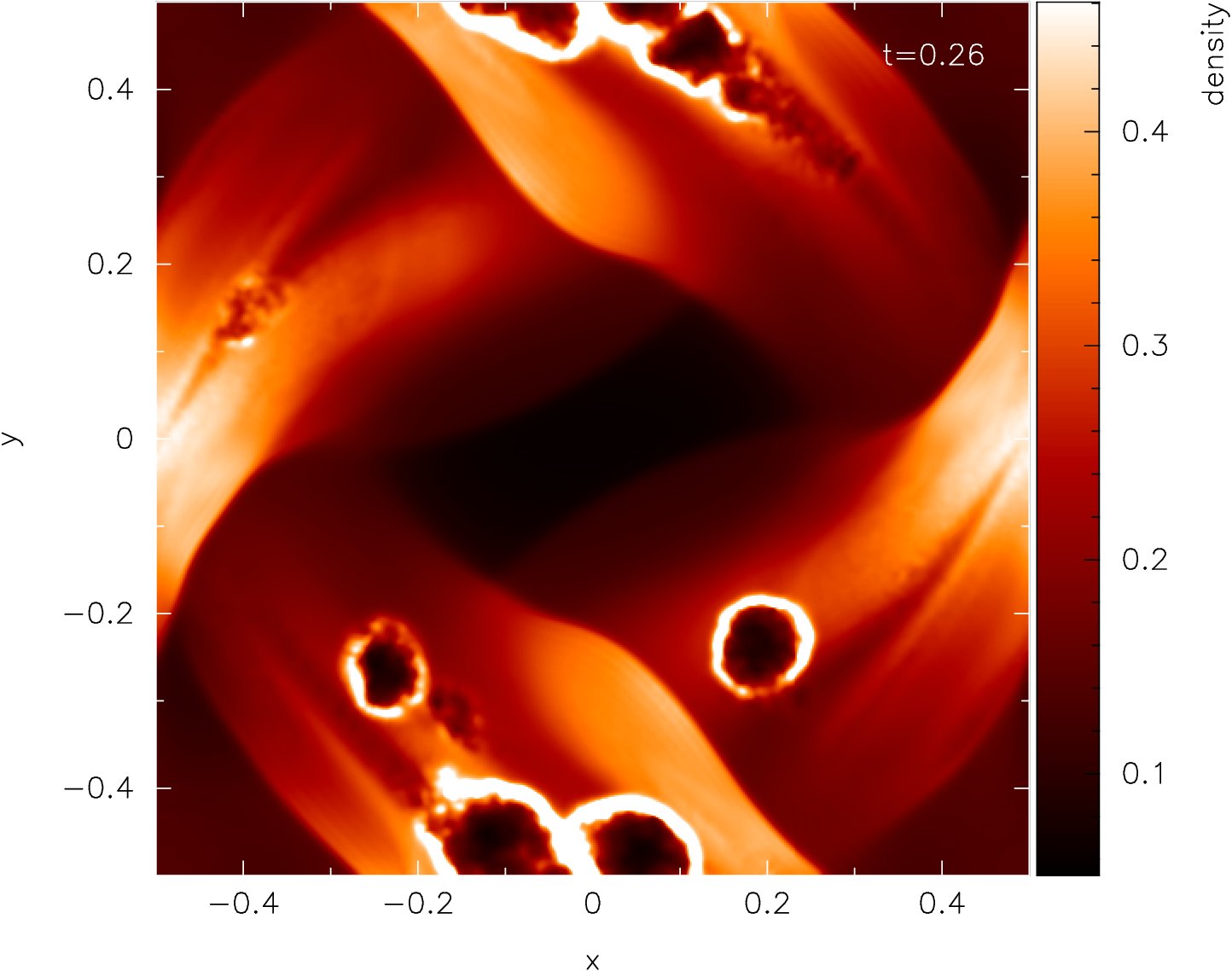}
\hspace{0.01\textwidth}
\includegraphics[angle=270,width=0.37\textwidth]{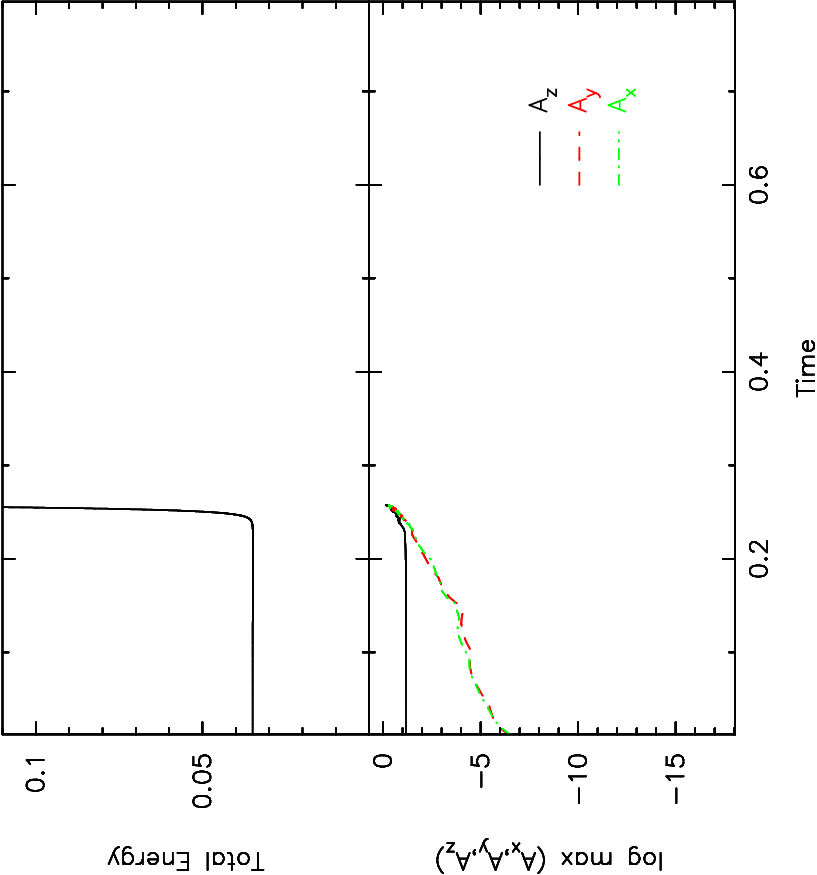}

\caption{Results of the Orszag-Tang Vortex evolution in 3D, using $128\times 128\times 16$ particles (top, using our {\sc ndspmhd} code) $100\times 100\times 10$ particles (middle, using {\sc phantom}) and $200\times 200\times 20$ particles (bottom, using {\sc phantom}). Here we have adopted the hybrid vector potential formulation that is stable to clumping instabilities and gives good results in 2D. In 3D we observe exponential growth of the $A_{x}$ and $A_{y}$ components of the vector potential (right panel, showing the maximum value as a function of time, alongside the evolution of total energy). When these components grow to the same order of magnitude as $A_{z}$, large, low density voids appear in the solution (left panels), together with an exponential divergence in total energy (right panels).}
\label{fig:orstang3D}
\end{figure*}

 Given that the consistent (energy conserving) vector potential formulation is unstable to the clumping instability for the Orzsag-Tang Vortex problem (\S\ref{sec:orstang}) --- which we expect (and find) is equally true 3D as much as it is in 2D --- we consider only the hybrid formulation in 3D, by which we mean using equation (\ref{eq:botforce}) for the force instead of (\ref{eq:force}). The advantage of a 3D problem is that the evolution equation for the vector potential, rather than simply being $dA_{z}/dt = 0$ in 2D, is given by (\ref{eq:dAdt}), implemented as (\ref{eq:dAdtSPH}) which also differs considerably from the evolution equations for the Euler potentials (\ref{eq:eulerevol}).
 
 In 3D we set up the problem similarly to \citet{ds08}, by adding a shortened $z$ dimension to the 2D box, placing particles initially on  a cubic lattice in the domain $x,y \in [-0.5,0.5]$ and $z \in [-0.0625,0.0625]$. The setup parameters are identical to the 2D problem described in \S\ref{sec:orstang} apart from the 3D domain. We have first computed the problem using the same {\sc ndspmhd} code that we have used above for the 2D problems, using zero artificial resistivity as in the 2D solution shown in \citet{rp07}. Given our findings below, we have also computed the solution as a consistency check using the implementation of the hybrid scheme in our {\sc phantom} SPH code (used in \citet{kitsionasetal09} and \citet{pf09}), which being parallelised, was also used to compute a higher resolution version. We show results in Figure~\ref{fig:orstang3D} using $128\times 128\times 16$ particles with {\sc ndspmhd} (top row), $100\times 100\times 10$ particles using {\sc phantom} (middle row) and $200\times 200\times 20$ particles (bottom row), also using {\sc phantom}.
 
 The results initially (not shown) are similar to the 2D results --- and therefore quite reasonable --- but only for a finite time. We find that in the 3D case the initially zero $A_{x}$ and $A_{y}$ components of the magnetic field grow exponentially with time, illustrated in the right panels of Figure~\ref{fig:orstang3D} which show the maximum (absolute) value of the $A_{x}$, $A_{y}$ and $A_{z}$ components of the vector potential as a function of time (bottom panel in the right hand figures), alongside the evolution of the total energy (top panel in the right hand figures). The growth of these unphysical components is initiated by simple round-off error, evident from the differing starting values between the {\sc phantom} results (bottom two rows) which stores the velocity and magnetic fields as 4-byte variables (giving a starting value of  max$(A_{x}, A_{y}) \sim 10^{-8}$), compared with the {\sc ndspmhd} results (top row), which stores both fields to 8-byte precision, giving a starting value of max$(A_{x}, A_{y}) \sim 10^{-16}$ similar to the round-off error in (e.g.) momentum conservation (Figure~\ref{fig:orstangen}) using double precision variables. The growth \emph{rate} is exponential in all cases, with max$(A_{x}, A_{y}) \propto \sim e^{20 t}$ for the top and bottom rows in Figure~\ref{fig:orstang3D} and max$(A_{x}, A_{y}) \propto \sim e^{10 t}$ for the middle row.
 
   At the point where the $A_{x}$ and $A_{y}$ components become similar in magnitude to the (physical) $A_{z}$ component, large low density voids appear in the density field (left panels of Figure~\ref{fig:orstang3D}) and correspondingly exponential growth in the total energy is observed (right panels of Figure~\ref{fig:orstang3D}), bringing the simulation rapidly to a halt. We also find the same outcome, though with variations in the exact time at which the simulation is disrupted, regardless of whether the Galilean invariant gauge (\ref{eq:dAdttensor}) or the standard gauge (\ref{eq:standardgauge}) is used to evolve the vector potential. In general the exact nature of the disruption caused by the growth of unphysical components of the vector potential in 3D depends on small details such as round-off error in the code and the nature of the problem studied, however we find similar problems attempting to use the vector potential in 3D star formation problems. It should be noted that no such problems arise with the use of the Euler potentials for the 3D Orszag-Tang Vortex, since the $dA_{z}/dt = 0$ evolution corresponds to the $d\alpha/dt = 0$ part of (\ref{eq:eulerevol}) whilst the $\beta$ variable is initially set to the $z$ position of the particles, in which we observe no change, corresponding to the $d\beta/dt = 0$ part of (\ref{eq:eulerevol}).

   Actually the results shown in Figure~\ref{fig:orstang3D} bring our investigation full circle. In fact, we started our examination of the vector potential as an alternative to the Euler potentials by studying the 3D Orszag-Tang Vortex, finding the results discussed above. The hope was that a consistent formulation of the vector potential from a variational principle, that guarantees exact energy conservation in the code and furthermore directly couples both the gauge (\ref{eq:dAdt}) and the numerical formulation of $d{\bf A}/dt$ equation (\ref{eq:dAdtSPH}) to the force formulation (\ref{eq:force}) would resolve the instabilities observed in Figure~\ref{fig:orstang3D}. However, as has already been discussed, the consistent formulation turns out to be itself unstable to the tensile instability known to plague conservative formulations of standard SPMHD. The journey is summarised below.

\section{Discussion}
\label{sec:summary}
 In this paper we have considered the use of the vector potential as a representation for the magnetic field in the context of the Lagrangian Smoothed Particle Magnetohydrodynamics method. In particular, we have addressed the question of whether using the vector potential may resolve the issues relating to the restrictions placed on the evolution of 3D magnetic fields using the Euler potentials. To this end we have derived a consistent, Hamiltonian formulation for vector potential SPMHD that guarantees both the conservation of momentum and energy. The formulation itself relies only on the choice of SPH formulations for the density summation (\S\ref{sec:densitysum}), for obtaining the magnetic field from the vector potential via ${\bf B} = \nabla\times{\bf A}$ (\S\ref{sec:BcurlA}) and for the evolution equation for ${\bf A}$ (\S\ref{sec:dAdt}), the latter of which involved an appropriate choice of gauge, which we required to be Galilean invariant in order to obtain momentum conservation in the equations of motion.

 From these three simple definitions, for which we have used standard variable-smoothing-length SPH operators, we have shown in \S\ref{sec:variational} that the equations of motion can be derived self-consistently from a Lagrangian variational principle, resulting in a form that, reflecting the symmetries inherent in the Lagrangian and associated constraint equations, indeed conserves momentum, energy and entropy simultaneously. The result is an expression for the MHD (Lorentz) force in SPH that is unique to the vector potential and which differs from all previous SPMHD force formulations. The force, given by (\ref{eq:force}), or more compactly by (\ref{eq:equationsofmotionstress}) can be broken down into components that are non-zero either for different numbers of spatial dimensions or depending on whether or not an external magnetic field is applied.
 
  The expression for the force given by (\ref{eq:force}), particularly the ``2D'' component, initially gave us hope that it might, after all, have been possible to formulate equations of motion for SPMHD that are both conservative and stable with respect to the tensile or ``clumping'' instability \citep{monaghan00} that prevents the use of exactly conservative force formulations in the standard SPMHD approach \citep{bot04,pm04a,pm05} and was the cause of many initial problems with SPMHD \citep{pm85}. This was not an unreasonable expectation, since the tensile instability in standard SPMHD is caused by non-zero $\nabla\cdot{\bf B}$ terms in the equations of motion --- whereas for the vector potential the knowledge that $\nabla\cdot{\bf B} = 0$ can be ``built-in'' to the equations of motion which are derived from ${\bf B} = \nabla\times{\bf A}$. However it turns out that the 2D term is balanced by a highly unstable isotropic term, for which the net pressure is negative (the trigger point for the tensile instability) when $3/2 B^{2}/\mu_{0} > P$, that is, worse than for the standard SPMHD case and not even related solely to the anisotropic part of the force \citep{price04}. In numerical tests (\S\ref{sec:tests}, particularly \S\ref{sec:alfven} and \S\ref{sec:orstang}) we have found that the instability indeed sets in at around the above threshold. Whilst we find that adding correction terms to the force (\S\ref{sec:correct}) --- though immediately violating the conservation properties we so desired --- does provide stability for certain problems, we find that the degree of correction required by other problems (\S\ref{sec:orstang}) starts to modify the solution unphysically. In principle a full stability analysis of the consistent vector potential formulation (\ref{eq:force}) would at least yield insights into the exact regime of stability, though we are not confident that much would be gained from such an analysis, let alone a suitable correction.
  
  The form of the isotropic term causing the tensile instability in (\ref{eq:force}) arises primarily from the weighting with respect to $\rho$ inherent in our SPH expression for ${\bf B} = \nabla\times{\bf A}$, equation (\ref{eq:BcurlA}), for which we have used a standard curl operator \citep[e.g.][]{price04}. It is possible that re-formulating this equation using different weightings of $\rho$ (e.g., using $1/\rho_{b}^{2}$ inside the summation) would result in a stable formulation, and this would certainly be an approach worth pursuing. However, we also find problems with the vector potential evolution equation even using a standard SPMHD force (\S\ref{sec:orszagtang3D}), so it is not clear that solving the tensile-instability related problems with the consistent formulation would necessarily solve all of the problems.

 An issue not addressed in detail in this paper given the severity of other problems was the fact that the 2D term in (\ref{eq:force}) involves direct second derivatives of the kernel, which are known to be particularly poor using the cubic spline kernel (\ref{eq:cubic}) since it is discontinuous in the second derivative. We have side-stepped the issue in this paper by using the smoother quintic kernel (\ref{eq:quintic}) where appropriate, though remarkably we find quite reasonable results could be obtained in many cases (for example the 1D shock tube problems in \S\ref{sec:shocktubes}) even using the cubic spline. The issue of formulating second derivatives is well known in SPH \citep[see e.g.][]{brookshaw85,price04,monaghan05}, however it is fairly straightforward to show that the usual Brookshaw formulation is equivalent to simply choosing a more appropriate form of the kernel for use in the second derivative calculation --- that is, using the second derivative of a bell-shaped kernel appropriate to density estimates is perhaps not the best approach. For the vector potential we have the freedom to choose the kernel that enters the second derivative term, which arises from the kernel in (\ref{eq:curlA}), completely separately from the kernel used in the density sum. That is, with no loss of consistency, the kernels in (\ref{eq:curlA}) and (\ref{eq:rhosum}) do not have to be the same (we have merely assumed that they were in this paper for simplicity). Thus an investigation into the best form of the kernel for computing the curl operation (\ref{eq:curlA}) separate from the density sum, perhaps with the second derivative estimate (\ref{eq:force}) in mind, would be a worthwhile exercise.
 
 Perhaps the most useful aspect of this paper --- apart from acting as a warning to the reader intent on similar endeavours --- is the formulation of dissipative terms for the vector potential presented in \S\ref{sec:diss}. In particular it is clear from this section that the dissipative terms formulated by \citet{pb07} and \citet{rp07} for the Euler potentials were not correct, containing $\nabla\eta$ terms which should not be present, and more seriously not guaranteeing positive definite dissipation (though both those papers used a term in the thermal energy equation that \emph{is} positive definite, but which instead violates the exact conservation of energy). From an SPH algorithms perspective, \S\ref{sec:diss} very nicely illustrates the conjugate relationship between the standard (\ref{eq:curlA}) and symmetric (\ref{eq:symmetriccurl}) SPH curl operators\footnote{This has been noted previously by \citet{cr99}. Here we have demonstrated the equivalent relationship for the variable smoothing length SPH operators.}. Using the standard curl for ${\bf A}$ necessitates the use of the symmetric curl for ${\bf J}$ in order to obtain both energy conservation and positive definite dissipation. We have also shown in \S\ref{sec:diss} how the resistivity parameter $\eta$ may be constructed appropriate to an artificial resistivity term, demonstrating that this approach works well on standard shock tube problems (\S\ref{sec:shocktubes}) where dissipation is important.

 In terms of finding a satisfactory approach to maintaining the divergence constraint in three dimensional SPMHD simulations without the restrictions associated with the Euler Potentials formulation, we intend to examine generalised forms of the Euler Potentials which can represent arbitrary MHD fields and which can be more easily adapted for non-ideal MHD. This is deferred to a future paper.

\section{Conclusion}
 In summary, we find that using a hybrid formulation of the vector potential in SPH, evolving ${\bf A}$ using (\ref{eq:dAdtSPH}) and calculating the force using one of the standard stable SPMHD force expressions (\ref{eq:botforce}) or (\ref{eq:morrisforce}), is the approach with the fewest difficulties, and gives good results on one and two dimensional test problems -- identically to the Euler potentials in 2D. However, we also find problems with this approach for 3D problems, leading us to conclude that use of the vector potential is not a viable approach for SPMHD.

\section*{Acknowledgements} 
We thank Axel Brandenburg for urging development of the vector potential formulation in the first place and for suggesting the gauge choice, Matthew Bate and Joe Monaghan for useful discussions, and Giuseppe Lodato for a helpful referee's report. DJP acknowledges the support of a Monash fellowship. Figures were produced using SPLASH \citep{splashpaper}: http://users.monash.edu.au/$\sim$dprice/splash.
\appendix

\section{Perturbation to the magnetic field}
\label{sec:deltaBrho}
The derivation of the Lagrangian perturbation for the magnetic field, discussed in \S\ref{sec:perturb} and used in equation (\ref{eq:deltarhoB}), is easier to understand if one considers that the Lagrangian perturbation $\delta \equiv \Delta + \delta {\bf x}\cdot\nabla$ is similar to taking a Lagrangian time derivative $d/dt \equiv \partial/\partial t + {\bf v}\cdot\nabla$ (more precisely $d/dt \equiv \delta/\delta t$). Taking the time derivative of (\ref{eq:curlA}), assuming $d{\bf B}_{ext}/dt = 0$, gives
\begin{eqnarray}
\frac{d{\bf B}_{a}}{dt} &= & \frac{1}{\Omega_{a}\rho_{a}} \sum_{b} m_{b} \left({\bf A}_{a} - {\bf A}_{b} \right) \times \frac{d}{dt}\left[\nabla_{a} W_{ab} (h_{a})\right] \nonumber \\
& + & \frac{1}{\Omega_{a}\rho_{a}} \sum_{b} m_{b} \left(\frac{d{\bf A}_{a}}{dt} - \frac{d{\bf A}_{b}}{dt} \right) \times \nabla_{a} W_{ab} (h_{a}) - \frac{{\bf B}_{int}}{\rho_{a}} \frac{d\rho_{a}}{dt} -\frac{{\bf B}_{int}}{\Omega_{a}} \frac{d\Omega_{a}}{dt}.
\label{eq:appdBdt}
\end{eqnarray}
The time derivative of the kernel gradient can be shown (Appendix \ref{sec:dgradWdt}) to be given by
\begin{equation}
\frac{d}{dt} (\nabla_{a} W_{ab}) = ({\bf v}_{ab}\cdot\nabla) \nabla W_{ab}(h_{a}) + \pder{\nabla W_{ab}(h_{a})}{h_{a}} \frac{dh_{a}}{dt},
\label{eq:dgradWdt}
\end{equation}
the latter term arising only in the case of a variable smoothing length. Using this expression in (\ref{eq:appdBdt}) and assuming $h = h(\rho)$ gives
\begin{eqnarray}
\frac{d{\bf B}_{a}}{dt} &= & \frac{1}{\Omega_{a}\rho_{a}} \sum_{b} m_{b} \left({\bf A}_{a} - {\bf A}_{b} \right) \times [({\bf v}_{a} - {\bf v}_{b})\cdot\nabla] \nabla_{a} W_{ab} (h_{a}) \nonumber \\
& + & \frac{1}{\Omega_{a}\rho_{a}} \sum_{b} m_{b} \left(\frac{d{\bf A}_{a}}{dt} - \frac{d{\bf A}_{b}}{dt} \right) \times \nabla_{a} W_{ab} (h_{a})
 - \frac{{\bf B}_{int}}{\rho_{a}} \frac{d\rho_{a}}{dt} - \frac{{\bf B}_{int}}{\Omega_{a}} \frac{d\Omega_{a}}{dt}   +  \frac{\hterm_{a}}{\rho_{a}} \frac{d\rho_{a}}{dt},
\end{eqnarray}
where we have defined
\begin{equation}
\hterm_{a} \equiv \pder{h_{a}}{\rho_{a}} \frac{1}{\Omega_{a}} \sum_{b} m_{b} \left({\bf A}_{a} - {\bf A}_{b} \right) \times \pder{\nabla_{a} W_{ab} (h_{a})}{h_{a}}.
\label{eq:hsum}
\end{equation}
Replacing ${\bf v}$ with $\delta {\bf x} / \delta t$ and time derivatives $d/dt$ with $\delta/\delta t$ we have
\begin{eqnarray}
\delta {\bf B}_{a} &= & \frac{1}{\Omega_{a}\rho_{a}}\sum_{b} m_{b} ({\bf A}_{a} - {\bf A}_{b}) \times [(\delta {\bf x}_{a} - \delta {\bf x}_{b})\cdot\nabla] \nabla_{a} W_{ab} (h_{a})
 \nonumber \\
& + & \frac{1}{\Omega_{a}\rho_{a}} \sum_{b} m_{b} \left(\delta{\bf A}_{a} - \delta{\bf A}_{b} \right) \times \nabla_{a} W_{ab} (h_{a}) - \frac{{\bf B}_{int}}{\rho_{a}} \delta\rho_{a} 
 -\frac{{\bf B}_{int}}{\Omega_{a}} \delta\Omega_{a} 
+ \frac{\hterm_{a}}{\rho_{a}} \delta\rho_{a},
\label{eq:deltaB}
\end{eqnarray}
This is an SPH expression for the Lagrangian perturbation of the magnetic field, i.e.,
\begin{equation}
\delta{\bf B} = \nabla\times [\delta{\bf A} - (\delta{\bf x}\cdot\nabla){\bf A}] + \delta{\bf x}\cdot\nabla (\nabla\times{\bf A}).
\end{equation}

 The perturbation required in \S\ref{sec:perturb}, i.e., $\delta (\rho_{b}{\bf B}_{b}) = \rho_{b} \delta{\bf B}_{b} + \delta\rho_{b} {\bf B}_{b}$ is thus given by
\begin{eqnarray}
\delta (\rho_{b} {\bf B}_{b}) &= & \frac{1}{\Omega_{b}}\sum_{c} m_{c} ({\bf A}_{b} - {\bf A}_{c}) \times [(\delta {\bf x}_{b} - \delta {\bf x}_{c})\cdot\nabla] \nabla_{b} W_{bc} (h_{b}) \nonumber \\
& + & \frac{1}{\Omega_{b}} \sum_{c} m_{c} \left(\delta{\bf A}_{b} - \delta{\bf A}_{c} \right) \times \nabla_{b} W_{bc} (h_{b}) + {\bf B}_{ext}\delta\rho_{b} -  \frac{{\bf B}_{b,int}\rho_{b}}{\Omega_{b}} \delta\Omega_{b} + \hterm_{b} \delta\rho_{b} . \label{eq:deltaBrho}
\end{eqnarray}

 The higher order perturbations $\delta\Omega$ relating to the smoothing length gradients are of order the truncation error of the SPH method (i.e., $\mathcal{O}(h^{2})$) and thus may be justifiably neglected from the analysis. On the other hand, to obtain a method that conserves energy to timestepping accuracy it is necessary to include these terms. Thus, for completeness, the perturbation to $\Omega$, from Equation (\ref{eq:omega}), is given by
\begin{equation}
\delta\Omega_{a} = -\frac{\zeta_{a}}{\rho_{a}}\delta\rho_{a} - \pder{h_{a}}{\rho_{a}} \sum_{b} m_{b} \left[(\delta {\bf x}_{a} - \delta {\bf x}_{b})\cdot\nabla\right] \pder{W_{ab}(h_{a})}{h_{a}},
\label{eq:deltaomega}
\end{equation}
where we have defined the dimensionless quantity
\begin{equation}
\zeta_{a} \equiv \rho_{a}\left[\pder{^{2} h_{a}}{\rho_{a}^{2}} \sum_{b} m_{b} \pder{W_{ab}(h_{a})}{h_{a}} + \left(\pder{h_{a}}{\rho_{a}} \right)^{2} \sum_{b} m_{b} \pder{^{2} W_{ab}(h_{a})}{h_{a}^{2}}\right].
\label{eq:zeta}
\end{equation}

 Combining (\ref{eq:deltaomega}) and (\ref{eq:deltaBrho}), the full perturbation including all terms relating to the smoothing length gradients is given by
\begin{eqnarray}
\delta (\rho_{b} {\bf B}_{b}) &= & \frac{1}{\Omega_{b}}\sum_{c} m_{c} ({\bf A}_{b} - {\bf A}_{c}) \times [(\delta {\bf x}_{b} - \delta {\bf x}_{c})\cdot\nabla] \nabla_{b} W_{bc} (h_{b}) \nonumber \\
& + & \frac{1}{\Omega_{b}} \sum_{c} m_{c} \left(\delta{\bf A}_{b} - \delta{\bf A}_{c} \right) \times \nabla_{b} W_{bc} (h_{b}) + {\bf B}_{ext}\delta\rho_{b} \nonumber \\
& + & \left[ \hterm_{b} + \frac{{\bf B}_{b,int}}{\Omega_{b}} \zeta_{b} \right] \delta\rho_{b} + \frac{{\bf B}_{b,int}\rho_{b}}{\Omega_{b}}\pder{h_{b}}{\rho_{b}} \sum_{c} m_{c} \left[(\delta {\bf x}_{b} - \delta {\bf x}_{c})\cdot\nabla_{b}\right] \pder{W_{bc}(h_{b})}{h_{b}}.
\label{eq:deltaBrhofull}
\end{eqnarray}

\section{Kernel functions and derivatives}
\label{sec:dgradWdt}
\label{sec:kernelderivs}
\label{sec:kernel}
 In this appendix we describe the various kernel functions used in this paper and derive the derivatives required in the vector potential formulation, showing how they may be computed from the dimensionless kernel functions.

\subsection{Kernel function}
 It is common to express the SPH kernel in the form
\begin{equation}
W_{ab}(h) \equiv \frac{\sigma}{h^{\nu}} f(q),
\end{equation}
where $\sigma$ is a normalisation constant, $h$ is the smoothing length, $\nu$ is the number of spatial dimensions and $f(q)$ is a dimensionless function of the variable $q \equiv \vert {\bf r}_{a} - {\bf r}_{b} \vert / h$. For the standard SPH cubic spline kernel \citep[e.g.][]{monaghan92} the function $f$ is given by
\begin{equation}
f(q) = 
\begin{cases}
\frac14(2-q)^{3} - (1-q)^{3}, & \text{$0 \leq q < 1$;} \\
\frac14(2-q)^{3}, & \text{$1 \leq q < 2$;}\\
0,                & \text{$q \geq 2$}.
\end{cases}
\label{eq:cubic}
\end{equation}
with normalisation $\sigma = [2/3, 10/(7\pi), 1/\pi]$ in $[1,2,3]$ dimensions. In this paper we also use the quintic spline kernel given by \citep[e.g.][]{morrisphd}
\begin{equation}
f(q) = 
\begin{cases}
(3-q)^{5} - 6(2-q)^{5} + 15(1-q)^{5}, & \text{$0 \leq q < 1$;} \\
(3-q)^{5} - 6(2-q)^{5}, & \text{$1 \leq q < 2$;} \\
(3-q)^{5}, & \text{$2 \leq q < 3$;} \\
0,                & \text{$q \geq 3$}.
\end{cases}
\label{eq:quintic}
\end{equation}
with normalisation $\sigma = [1/24, 96/(1199\pi), 1/(20\pi)]$. The advantage of the quintic is that it more closely approximates the Gaussian by extending the compact support radius to $3h$ and has smooth second derivatives --- important here since we use the second derivative directly in the force (equation \ref{eq:force}). The disadvantage is that it is considerably more expensive to compute, particularly in three dimensions, where the neighbour number, and thus the cost, increases by a factor of $(3/2)^{3} \approx 3.4$.

\subsection{Kernel first derivatives}
 The derivative of the kernel with respect to the particle coordinates, holding the smoothing length constant is given by
\begin{equation}
\nabla_{a} W_{ab} = \frac{\sigma}{h^{\nu}} f'(q) \nabla q = \hat{\bf r}_{ab} \frac{\sigma}{h^{\nu+1}} f'(q),
\label{eq:gradW}
\end{equation}
where we have used the fact that $\nabla q = \nabla(\vert r_{ab} \vert/h) = \hat{\bf r}_{ab}/h$, defining ${\bf r}_{ab} \equiv {\bf r}_{a} - {\bf r}_{b}$. The kernel derivative can also be written in the form
\begin{equation}
\nabla_{a} W_{ab} = \hat{\bf r}_{ab} F_{ab}, \hspace{0.5cm}\textrm{where}\hspace{0.5cm} F_{ab} \equiv \frac{\sigma}{h^{\nu+1}} f'(q).
\end{equation}
Note that the definition we use for $F_{ab}$ differs slightly from that used by \citet{monaghan92} since we use the unit vector in our definition, reflecting the implementation in our code.

 The derivative of the kernel with respect to $h$ is given in terms of $f(q)$ by
\begin{equation}
\pder{W_{ab}}{h_{a}} = -\nu \frac{\sigma}{h^{\nu+1}} f(q) + \frac{\sigma}{h^{\nu}} f'(q) \pder{q}{h} = - \frac{\sigma}{h^{\nu+1}} \left[\nu f(q) + q f'(q)\right].
\end{equation}

\subsection{Second derivatives}
 The second derivative of the kernel with respect to particle coordinates is given by
\begin{eqnarray}
\nabla^{i}_{a} \nabla^{j}_{a} W_{ab} & = & \nabla_{a}^{i} \left[ \hat{r}^{j}_{ab} \frac{\sigma}{h^{\nu+1}} f'(q) \right], \nonumber \\
& = & \frac{\sigma}{h^{\nu + 2}} \left[ \hat{r}^{i}_{ab} \hat{r}^{j}_{ab} f''(q) + \left(\delta^{ij} -  \hat{r}^{i}_{ab} \hat{r}^{j}_{ab} \right) \frac{1}{q} f'(q) \right].
\end{eqnarray}
For the Laplacian this reduces to
\begin{equation}
\nabla^{2} W_{ab} = \frac{\sigma}{h^{\nu + 2}} \left[ f''(q) + \left(\nu -  1 \right) \frac{1}{q} f'(q) \right].
\end{equation}

The time derivative of the kernel gradient can be derived from equation (\ref{eq:gradW}), giving
\begin{equation}
\frac{d}{dt} (\nabla_{a} W_{ab}) = \frac{{\bf v}_{ab} ({\bf r}_{ab}\cdot \nabla W_{ab})}{\vert r_{ab} \vert^{2}} -  \frac{({\bf v}_{ab} \cdot \nabla W_{ab}) {\bf r}_{ab}}{\vert r_{ab} \vert^{2}} + \frac{({\bf v}_{ab} \cdot  {\bf r}_{ab})}{\vert r_{ab} \vert} \frac{\sigma f''(q)}{h^{\nu + 2}} {\bf r}_{ab} \equiv ({\bf v}_{ab} \cdot\nabla) \nabla W_{ab}.
\end{equation}
 The mixed second derivative of the kernel with respect to particle coordinates and $h$ is given by
\begin{eqnarray}
\pder{}{h_{a}} \left(\nabla_{a} W_{ab}\right) & = & -(\nu + 1) \frac{\sigma}{h^{\nu + 2}} f'(q) \hat{\bf r}_{ab} + \frac{\sigma}{h^{\nu + 1}} f''(q) \pder{q}{h} \hat{\bf r}_{ab}, \nonumber \\
& = & -\frac{(\nu + 1)}{h} F_{ab} \hat{\bf r}_{ab} - \frac{\sigma}{h^{\nu + 2}} q f''(q) \hat{\bf r}_{ab}.
\end{eqnarray}
Finally, the second derivative with respect to the smoothing length is given by
\begin{eqnarray}
\pder{^{2} W_{ab}}{h^{2}_{a}} & = & \nu(\nu + 1) \frac{\sigma}{h^{\nu + 2}} f(q) + \frac{\sigma}{h^{\nu + 1}} \left[ \nu + (\nu + 2) \right] \frac{q}{h} f'(q) + \frac{\sigma q^{2}}{h^{\nu + 2}} f''(q), \nonumber \\
& = & \frac{\nu (\nu + 1)}{h^{2}} W_{ab} + \frac{2(\nu + 1) q}{h} F_{ab} + \frac{\sigma q^{2}}{h^{\nu + 2}} f''(q).
\end{eqnarray}

\section{Translation of the SPH equations of motion in the continuum limit}
\label{sec:translate}
 The MHD equations of motion are given by
\begin{equation}
\frac{d{\bf v}}{dt} = -\frac{\nabla P}{\rho} + \frac{{\bf J}\times{\bf B}}{\rho}.
\label{eq:JcrossB}
\end{equation}
 In this appendix we show that the SPH equations of motion derived in \S\ref{sec:equationsofmotion} translate to the above in the continuum limit. Written in terms of the vector potential, (\ref{eq:JcrossB}) becomes
\begin{equation}
\frac{d{\bf v}}{dt} = -\frac{\nabla P}{\rho} + \frac{{\bf J}\times (\nabla \times{\bf A})}{\rho} + \frac{{\bf J}\times {\bf B}_{ext}}{\rho},
\end{equation}
or, in tensor notation,
\begin{equation} 
\frac{dv^{i}}{dt} = -\frac{1}{\rho}\pder{P}{x^{i}} +\frac{1}{\rho}\left[J^{j}\pder{A^{j}}{x^{i}} - J^{j} \pder{A^{i}}{x^{j}} \right]  + \epsilon_{ijk} J^{j} B_{ext}^{k}.
\label{eq:mhdforceA}
\end{equation}

The terms in the SPH equations of motion (\ref{eq:force}), (\ref{eq:equationsofmotionstress}) or (\ref{eq:eomfixedh}) can be decoded into continuum form using the SPH summation interpolant
\begin{equation}
\sum_{b} m_{b} \frac{A_{b}}{\rho_{b}} W_{ab} \simeq A_{a},
\end{equation}
the derivatives of which give
\begin{equation}
 \sum_{b} m_{b} \frac{A_{b}}{\rho_{b}} \pder{W_{ab}}{x^{i}_{a}} \simeq \pder{A_{a}}{x^{i}_{a}}; 
 \hspace{1cm}
 \sum_{b} m_{b} \frac{A_{b}}{\rho_{b}} \pder{^{2} W_{ab}}{x^{i}_{a}x^{j}_{a}} \simeq \pder{^{2} A_{a}}{x^{i}_{a} x^{j}_{a}}.
 \label{eq:summationinterp}
\end{equation}
In translating the equations of motion we firstly neglect all terms relating to gradients in the smoothing length -- that is, we provide the translation of (\ref{eq:eomfixedh}) rather than (\ref{eq:force}). Whilst in principle the extra terms in (\ref{eq:force}) could also be translated, the proof that they are correctly derived lies in the fact that the numerical equations demonstrate the conservation of energy to timestepping accuracy, which we have shown in \S\ref{sec:tests}.
For a constant smoothing length, the SPH equations of motion (\ref{eq:eomfixedh}) can be written in tensor notation (akin to equation~\ref{eq:equationsofmotionstress}) in the form
\begin{equation}
\frac{dv^{i}_{a}}{dt} = \sum_{b} m_{b} \left[ \frac{ \mathcal{S}^{ij}_{a}}{\rho_{a}^{2}}  + \frac{ \mathcal{S}^{ij}_{b}}{\rho_{b}^{2}} \right] \pder{W_{ab}}{x^{j}_{a}} -  \frac{\epsilon_{jkl}}{\mu_{0}} \sum_{b} m_{b} (A_{k}^{a} - A_{k}^{b}) \left[ \frac{B^{j}_{a}}{\rho_{a}^{2}} + \frac{B^{j}_{b}}{\rho_{b}^{2}} \right] \pder{^{2} W_{ab}}{x^{i}_{a} \partial x^{l}_{a}},
\label{eq:eomfixedhstress}
\end{equation}
where
\begin{equation}
\mathcal{S}^{ij} = -P \delta^{ij} + \frac{1}{\mu_{0}}\left[B^{i}B^{j}_{ext} + \delta^{ij}\left(\frac32 B^{2}  - 2{\bf B}\cdot{\bf B}_{ext} \right) -  A^{i} J^{j} \right].
\label{eq:stressfixedh}
\end{equation}

 The first term in (\ref{eq:eomfixedhstress}) is similar to the usual conservative SPMHD force \citep[e.g.][]{pm04b}, albeit with a different stress tensor, and can be straightforwardly translated to $1/\rho$ $\partial \mathcal{S}^{ij} / \partial x^{j}$. Expanding this using (\ref{eq:stressfixedh}), we have
\begin{equation}
\frac{dv^{i}}{dt} = \frac{1}{\rho} \left[ -\pder{P}{x^{i}} + B_{ext}^{j} \pder{B^{i}}{x^{j}} + 3 B^{j} \pder{B^{j}}{x^{i}} - 2 B_{ext}^{j} \pder{B^{j}}{x^{i}} - J^{j} \pder{A^{i}}{x^{j}} - A^{i} \pder{J^{j}}{x^{j}}\right] +  \textrm{2nd term}
\label{eq:translation1}
\end{equation}
 The second or ``2D term'' in (\ref{eq:eomfixedhstress}) can be translated using (\ref{eq:summationinterp}) as follows
\begin{equation}
\textrm{2nd term} = -\frac{\epsilon_{jkl}}{\mu_{0}} A^{k} \frac{B^{j}}{\rho^{2}} \pder{^{2} \rho}{x^{i} \partial x^{l}} 
-\frac{\epsilon_{jkl}}{\mu_{0}} A^{k} \pder{}{x^{i}}\left[\pder{}{x^{l}} \left(\frac{B^{j}}{\rho} \right) \right]
-\frac{\epsilon_{jkl}}{\mu_{0}} \frac{B^{j}}{\rho^{2}}  \pder{^{2}(\rho A^{k})}{x^{i} \partial x^{l}}
+\frac{\epsilon_{jkl}}{\mu_{0}}\pder{}{x^{i}}\left[\pder{}{x^{l}} \left( \frac{A^{k} B^{j}}{\rho} \right)\right],
\end{equation}
where in the third term we obtain a change of sign because the indices $j$ and $k$ have changed order with respect to the Levi-Civita tensor. Expanding the derivative terms and collecting/cancelling terms where appropriate, after some straightforward algebra we are left with
\begin{equation}
- 2 \frac{\epsilon_{jkl}}{\mu_{0}} \frac{B^{j}}{\rho} \pder{^{2} A^{k}}{x^{i} x^{l}} +  \frac{\epsilon_{jkl}}{\mu_{0}\rho} \pder{A^{k}}{x^{l}} \pder{B^{j}}{x^{i}} +  \frac{\epsilon_{jkl}}{\mu_{0}\rho} \pder{A^{k}}{x^{i}} \pder{B^{j}}{x^{l}},
\end{equation}
which, using $B^{j}_{int} \equiv \epsilon_{jkl} \partial A^{k} / \partial x^{l}$ and $J^{k} \equiv 1/\mu_{0} \epsilon_{kjl} \partial B^{j} / \partial x^{l}$ gives
\begin{equation}
-2 \frac{B^{j}}{\mu_{0}\rho} \pder{B^{j}_{int}}{x^{i}} - \frac{B^{j}_{int}}{\mu_{0}\rho} \pder{B^{j}}{x^{i}} + \frac{1}{\rho}\pder{A^{k}}{x^{i}} J^{k}.
\end{equation}
Putting this together with (\ref{eq:translation1}) and collecting terms, noting that $B^{j} = B^{j}_{int} + B^{j}_{ext}$ and that we have previously assumed $\partial B^{j}_{ext}/\partial x^{i} = 0$ we have
\begin{equation}
\frac{dv^{i}}{dt} = \frac{1}{\rho} \left[ -\pder{P}{x^{i}} + \frac{B_{ext}^{j}}{\mu_{0}} \left( \pder{B^{i}}{x^{j}} - \pder{B^{j}}{x^{i}}\right) + \pder{A^{j}}{x^{i}} J^{j} - J^{j} \pder{A^{i}}{x^{j}} - A^{i} \pder{J^{j}}{x^{j}} \right].
\end{equation}
The last term is zero in the continuum limit since it is the divergence of a curl $(\nabla\cdot{\bf J}) \equiv \nabla\cdot(\nabla\times{\bf B})/\mu_{0}$. This completes the proof, giving the equations of motion in the form
\begin{equation}
\frac{dv^{i}}{dt} = \frac{1}{\rho} \left[ -\pder{P}{x^{i}} + \frac{B_{ext}^{j}}{\mu_{0}} \left( \pder{B^{i}}{x^{j}} - \pder{B^{j}}{x^{i}}\right) + \pder{A^{j}}{x^{i}} J^{j} - J^{j} \pder{A^{i}}{x^{j}} \right],
\end{equation}
which, upon expanding $\left[(\nabla\times{\bf B}) \times {\bf B}_{ext}\right]^{i} \equiv B_{ext}^{j} (\partial B^{i}/ \partial x^{j} - \partial B^{j}/ \partial x^{i})$, is identical to (\ref{eq:mhdforceA}).

\bibliography{sph,mhd,starformation}

\label{lastpage}
\enddocument